\documentclass[aps, pra,floatfix,superscriptaddress,showpacs, twocolumn, longbibliography, nofootinbib]{revtex4-1}
\usepackage{mathrsfs, amsmath, amsfonts, graphics, graphicx, cancel, color, theorem, bbm, mathtools, amssymb}

\usepackage[english]{babel}
\usepackage[margin=0.75in]{geometry}
\usepackage[T1]{fontenc}
\usepackage{xfrac,relsize,float}
\usepackage[unicode=true, breaklinks=false, pdfborder={0 0 1}, backref=false, colorlinks=true, linkcolor=blue, citecolor=blue, urlcolor=blue]{hyperref}
\usepackage{physics}
\usepackage{enumitem}
\usepackage{changepage}

\definecolor{Blue}{rgb}{0,0,1}
\definecolor{Red}{rgb}{1,0,0}
\definecolor{Green}{rgb}{0,1,0}
\definecolor{darkgreen}{rgb}{0,.7,0}
\definecolor{Purp}{rgb}{.2,0,.2}
\definecolor{white}{rgb}{1,1,1}

\usepackage{mathtools}

\DeclareMathAlphabet\mathbfcal{OMS}{cmsy}{b}{n}
\usepackage[most]{tcolorbox}

\newcommand{\eref}[1]{(\ref{#1})}
\newcommand{\eqnref}[1]{Eq.~(\ref{#1})}

\newcommand{\figref}[1]{Fig.~\ref{#1}}

\begin{document}
\title{ Bayesian quantum thermometry based on thermodynamic length }

\author{Mathias R. J{\o}rgensen}
    \email{matrj@fysik.dtu.dk}
    \affiliation{Department of Physics, Technical University of Denmark, 2800 Kongens Lyngby, Denmark}

\author{Jan Ko\l{}ody\'{n}ski}
    %\email{}
    \affiliation{Centre for Quantum Optical Technologies, Centre of New Technologies, University of Warsaw, 02-097 Warsaw, Poland}

\author{Mohammad Mehboudi}
    %\email{}
    \affiliation{D\'{e}partement de Physique Appliqu\'{e}e, Universit\'{e} de Gen\`{e}ve, 1211 Geneva, Switzerland}

\author{Mart\'{i} Perarnau-Llobet}
    %\email{}
    \affiliation{D\'{e}partement de Physique Appliqu\'{e}e, Universit\'{e} de Gen\`{e}ve, 1211 Geneva, Switzerland}

\author{Jonatan B. Brask}
    %\email{jonatan.brask@fysik.dtu.dk}
    \affiliation{Department of Physics, Technical University of Denmark, 2800 Kongens Lyngby, Denmark}

\date{\today}
\begin{abstract}
In this work, we propose a theory of temperature estimation of quantum systems, which is applicable in the regime of non-negligible prior temperature uncertainty and limited measurement data.
In this regime the problem of establishing a well-defined measure of estimation precision becomes non-trivial,
and furthermore the construction of a suitable criterion for optimal measurement design must be re-examined to account for the prior uncertainty.
We propose a fully Bayesian approach to temperature estimation based on the concept of thermodynamic length, which solves both these problems.
As an illustration of this framework, we consider thermal spin-$1/2$ particles and investigate the fundamental difference between two cases; on the one hand, when the spins are probing the temperature of a heat reservoir and, on the other, when the spins themselves constitute the sample.
\end{abstract}
\maketitle

%%%%%%%%%%%%%%%%%%%%%%%%%%%%%%%%%%%%%%%%%%%%%%%%%%%%%%%%%%%%%%%%%%%%%%%%%%%%%%%%%%%%%%%%%
%%%%%%%%%%%%%%%%%%%%%%%%%%%%%%%%%%%%%%%%%%%%%%%%%%%%%%%%%%%%%%%%%%%%%%%%%%%%%%%%%%%%%%%%%
\section{Introduction}
Measuring the temperature of a physical system is a fundamental task in science and technology.
At the micro- and nanoscale in particular, highly precise temperature measurements are essential for a large number of current experiments.
Examples include real-time monitoring of temperature profiles within living organisms e.g.,~utilizing colour centers in nanodiamonds~\cite{Kucsko_2013,Fujiwaraeaba_2020_inVivo,Moreva_2020},
the preparation of ultracold atoms in optical lattices, as well as mapping thermodynamic phase diagrams
and exploring transport phenomena~\cite{Carcy_2021,McKay_2011,Mitchison_2020_InSitu,Mehboudi_2019,Hartke_2020,Brantut_2013,Bouton_2020},
and studies of quantum thermodynamic phenomena in microelectronic devices~\cite{Gasparinetti_2015,Mecklenburg_2015,Halbertal_2016,Giazotto_2006,Karimi_2020_resolution}. Temperature is not a directly measurable property of a system, and in contrast to e.g.~interferometry, phase estimation, or electromagnetic-field sensing~\cite{Giovannetti_2011,Giovannetti_2006}, thermometry is further complicated by the fact that temperature is also not a Hamiltonian-encoded parameter.
Rather, the temperature of a system is an entropic quantity which must be estimated indirectly from the statistical behaviour of a variable which can be observed directly.
The purpose of the theory of quantum thermometry is both to guide the design of optimal measurement processes, i.e.,~building good thermometers in the quantum regime,
and to optimally infer from the acquired measurement data the underlying temperature~\cite{Mehboudi_2019_review,Pasquale_2018}.

The majority of previous works on quantum thermometric theory, with the notable exception of the recent studies~\cite{Rubio_2020_global,Alves_2021,Mok_2020},
have focused on local point estimation~\cite{Lehmann1998} -- termed for short the \emph{local paradigm} -- in which measurements are designed to detect small variations around a known temperature value~\cite{Mehboudi_2019_review,Pasquale_2018}.
Within the local paradigm, the expected precision of a temperature estimate is typically quantified by the frequentist mean-square error,
with the associated signal-to-noise ratio providing a meaningful notion of relative error~\cite{Kay1993}.
Given that certain conditions are satisfied, e.g.,~that the temperature estimate is \textit{unbiased},
the frequentist mean-square error is lower bounded, and typically well approximated, by the so-called Cram\'{e}r-Rao bound~\cite{Paris_2009,Bacharach_2019_BCRB,Braunstein_1994}.
Furthermore, optimal measurements applicable in the asymptotic (large data-set) regime can be identified via the local optimization of the Cram\'{e}r-Rao bound.

Motivation for constructing a theory applicable beyond the local paradigm is twofold:~(i) it is typically an unjustified assumption that the temperature to be estimated is known with sufficient precision a priori to justify working within the local paradigm, and (ii) the optimal measurement protocol generally depends on the prior temperature information, and cannot be identified via an optimization of the ``asymptotic'' Cram\'{e}r-Rao bound. Avoiding the restrictions of the local paradigm, i.e.,~providing a general approach to quantifying thermometric performance, and designing optimal measurements, under conditions of non-negligible prior uncertainty, requires a fully Bayesian framework~\cite{Udo_2011,van2007bayesian}.

\begin{figure}
    \centering
    \includegraphics[width=0.45\textwidth]{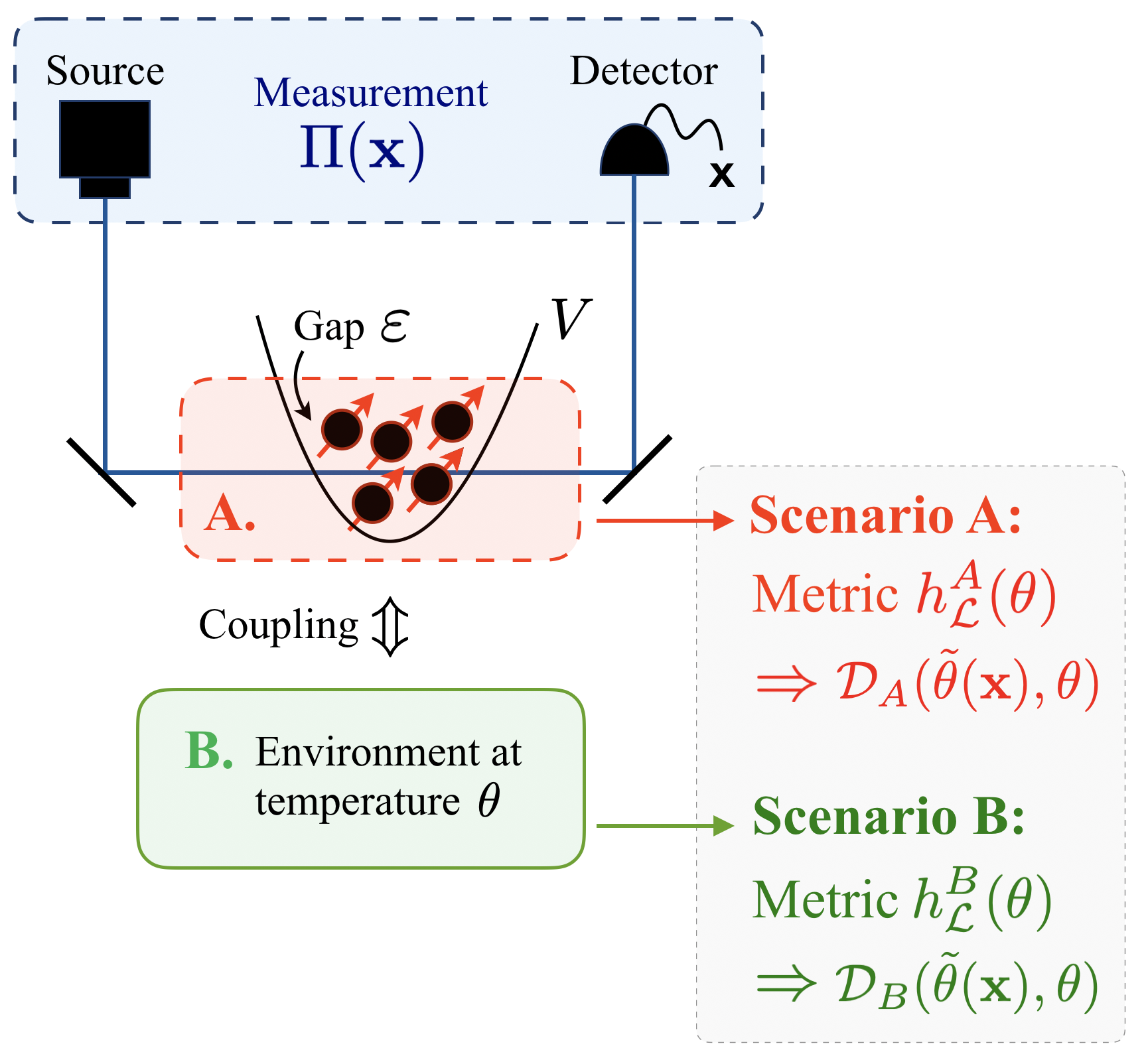}
    \caption{Illustration of a hypothetical experiment:
    A collection of spin impurities (A) interact with a thermal environment (B) at temperature $\theta$,
    the spin impurities are measured via an optical setup yielding measurement data $\textbf{x}$.
    An essential message of this paper is that the sample system defines the appropriate distance function, and that the distance is the basis of the estimation theory.
    To illustrate this point we consider two scenarios:
    (A)~The spin impurities themselves constitute the sample-system of interest,
    e.g.,~we are interested in thermometry of the spin degrees of freedom themselves~\cite{Hartke_2020,Weld_2009_SpinGradient},
    in this case we obtain a metric $h_\mathcal{L}^{A}$ and a distance $\mathcal{D}_{A}$.
    (B)~The spin impurities are employed as equilibrium thermometers of the thermal environment~\cite{Bouton_2020,Olf_2015_BEC},
    in this case we obtain a different metric $h_\mathcal{L}^{B}$ and a different distance $\mathcal{D}_{B}$.
    As a consequence, the relevant figure of merit, and associated optimal protocol, will be different in each sensing process.}
    \label{fig:conceptual_illustration}
\end{figure}

In this work we develop a theory of Bayesian quantum thermometry, applicable for any amount of prior information,
which is based on the concept of a thermodynamic length~\cite{Weinhold1975,Salamon1983,Crooks_2007_thermo_length,Scandi_2019_thermodynamiclength}.
The basic idea is that a meaningful measure of thermometric precision should be based on the ability to distinguish states at different temperatures,
i.e., colder from hotter, and should be independent of the particular parameterization of the states, e.g., temperature.
This can be naturally achieved by introducing a distance function between the thermal states of the sample system considered.
Such a distance is exactly the thermodynamic length between thermal states~\cite{Weinhold1975,Salamon1983,Crooks_2007_thermo_length,Scandi_2019_thermodynamiclength},
and we argue that this choice is singled out by the requirement that a well-defined distance should respect the invariance properties of the sample.
An interesting implication of the proposed framework is that any meaningful definition of relative error, must be given with respect to the specific sample system considered.
This feature is illustrated in Fig.~\ref{fig:conceptual_illustration}.
In particular, we find that the standard noise-to-signal ratio, defined in terms of the frequentist mean-square error,
is only recovered as a meaningful relative error -- within the local regime -- when the considered sample system can be effectively modelled as an ideal heat bath.

For the sake of illustration, we consider a thermometry setting that involves non-interacting spin-$1/2$ particles,
and compare the scenario in which the interest is in thermometry of the particles themselves, to the scenario in which the particles are employed as probes of an underlying heat reservoir.
Specifically, we look at simulated outcomes of projective energy measurements of thermal spin-$1/2$ particles, and compare the computed temperature estimates and measures of precision in the two cases.
The example illustrates that the rate of convergence to the local regime, and the suitability of various precision bounds, depends on the specific scenario considered.

%%%%%%%%%%%%%%%%%%%%%%%%%%%%%%%%%%%%%%%%%%%%%%%%%%%%%%%%%%%%%%%%%%%%%%%%%%%%%%%%%%%%%%%%%
%%%%%%%%%%%%%%%%%%%%%%%%%%%%%%%%%%%%%%%%%%%%%%%%%%%%%%%%%%%%%%%%%%%%%%%%%%%%%%%%%%%%%%%%%
\section{Bayesian estimation theory}
In this section we summarize the key concepts of Bayesian estimation theory~\cite{van2007bayesian,Udo_2011}.
The aim of estimation theory, at least when considering point estimation~\cite{Lehmann1998}, is to provide a prescription for singling out an estimate, serving as a best guess of an unknown quantity of interest,
and for quantifying a measure of confidence in, or the precision of, the specified estimate. In this work, we specifically consider a smooth manifold of quantum states, e.g., thermal Gibbs states, of a sample---depicted schematically as $\mathcal{S}$ in \figref{fig:illustration}. Our task is to estimate the true state of the sample, assuming that this belongs to the specified manifold.

Within the Bayesian theory, the manifold of states is equipped with a probability distribution, and this distribution is updated as measurement data is acquired.
An estimate is computed, according to some prescription, based on the updated probability distribution. Providing a measure of confidence in an estimate, requires a notion of distance between states on the manifold. Here, we first argue for a metric structure on the manifold of states. A distance function can then be constructed as the geodesic length between states.
As a result, the confidence in the computed estimate can then be unambiguously quantified via the mean-square distance over the updated probability distribution.

Having established the required notions of estimation theory, we provide a criterion for the optimal measurement design,
and show how the local estimation theory can be recovered as the asymptotic limit of the Bayesian framework.
Moreover, we outline a number of known results on Bayesian Cram\'{e}r-Rao bounds that are particularly useful.
Lastly, we discuss how to select an uninformative initial prior probability distribution.

\subsection{Manifold of states and Bayesian updating}
Physical systems are typically subject to a set of experimental conditions, e.g.,~some preparation procedure,
which defines a smooth manifold of states~\cite{Gallavotti_1999,Guarnieri_2019,Brenes_2020}.
To be specific we consider a quantum system, the \textit{sample},
and assume that the true, but unknown, state of the sample~$\rho_{\text{true}}$, belongs to a one-parameter family~$\mathcal{S}_\Theta$ of quantum states in the manifold
\begin{equation}
    \rho_{\text{true}} \in \mathcal{S}_\Theta \coloneqq \left\lbrace \tau_{\Theta}(\theta) \ \ \text{for} \ \ \theta\in \Theta \right\rbrace ,
\end{equation}
where $\tau_{\Theta}(\theta)$ labels a parameterized quantum state, i.e.,~a linear operator on the sample Hilbert space,
and $\Theta \subseteq \mathbb{R}$ denotes the parameter space, e.g.,~$\Theta$ could be the space of temperatures.
Our task is to identify the true sample state, which, given that the state belongs to a one-parameter family, can be formulated as a parameter estimation problem.
Throughout, we focus on a one-dimensional parameter space, and leave the generalization to the multi-dimensional case for future work.

Within Bayesian estimation theory we start from a prior probability density $p(\theta)$ on the parameter space $\Theta$,
which is subsequently updated as the sample state is probed via measurements.
Quantum mechanically, any measurement can be represented by a positive operator-valued measure ($\small{\text{POVM}}$).
If we suppose that the sample is in the state $\tau_{\Theta}(\theta)$, then the likelihood $p(x\!\mid\! \theta)$ of observing a measurement outcome $x$ is given by Born's rule
\begin{equation}
    p(x\!\mid\! \theta) = \Tr\left[ \Pi(x) \tau_{\Theta}(\theta) \right] ,
\end{equation}
where $\Pi(x)$ is the $\small{\text{POVM}}$ element associated with the outcome $x\in X$, and $X$ is the outcome space of the measurement performed.
The $\small{\text{POVM}}$ elements must be positive semi-definite, i.e. $\Pi(x)\geqslant 0$, and satisfy the normalization condition $\int_{X}dx \Pi(x) = \mathbb{I}$ with $\mathbb{I}$ being the identity operator~\cite{Ballentine_2014_book}.
Conditioned on observing the specific outcome $x$, we can find the posterior probability distribution using Bayes' rule:
\begin{equation}
    p(\theta \!\mid\! x) = \frac{p(x\!\mid\! \theta) p(\theta)}{p(x)} ,
\end{equation}
where $p(x) := \int_{\Theta} d\theta p(x\!\mid\! \theta) p(\theta)$ is the marginal probability density on the space of outcomes.
For later use we also define the joint density $p(\theta,x) =p(x\!\mid\! \theta) p(\theta)$.
The posterior density represents our degree of belief in different parameter values given the available measurement data.
Note that here we formulate Bayes' rule as a single-shot update. If we consider $\nu$ independent measurements giving outcomes $\textbf{x}=\left\lbrace x_{1},...,x_{\nu} \right\rbrace$,
then it is equally valid to compute the posterior conditioned on the full set of observations.

\subsection{Parameterization invariance}
\begin{figure}
    \centering
    \includegraphics[width=0.45\textwidth]{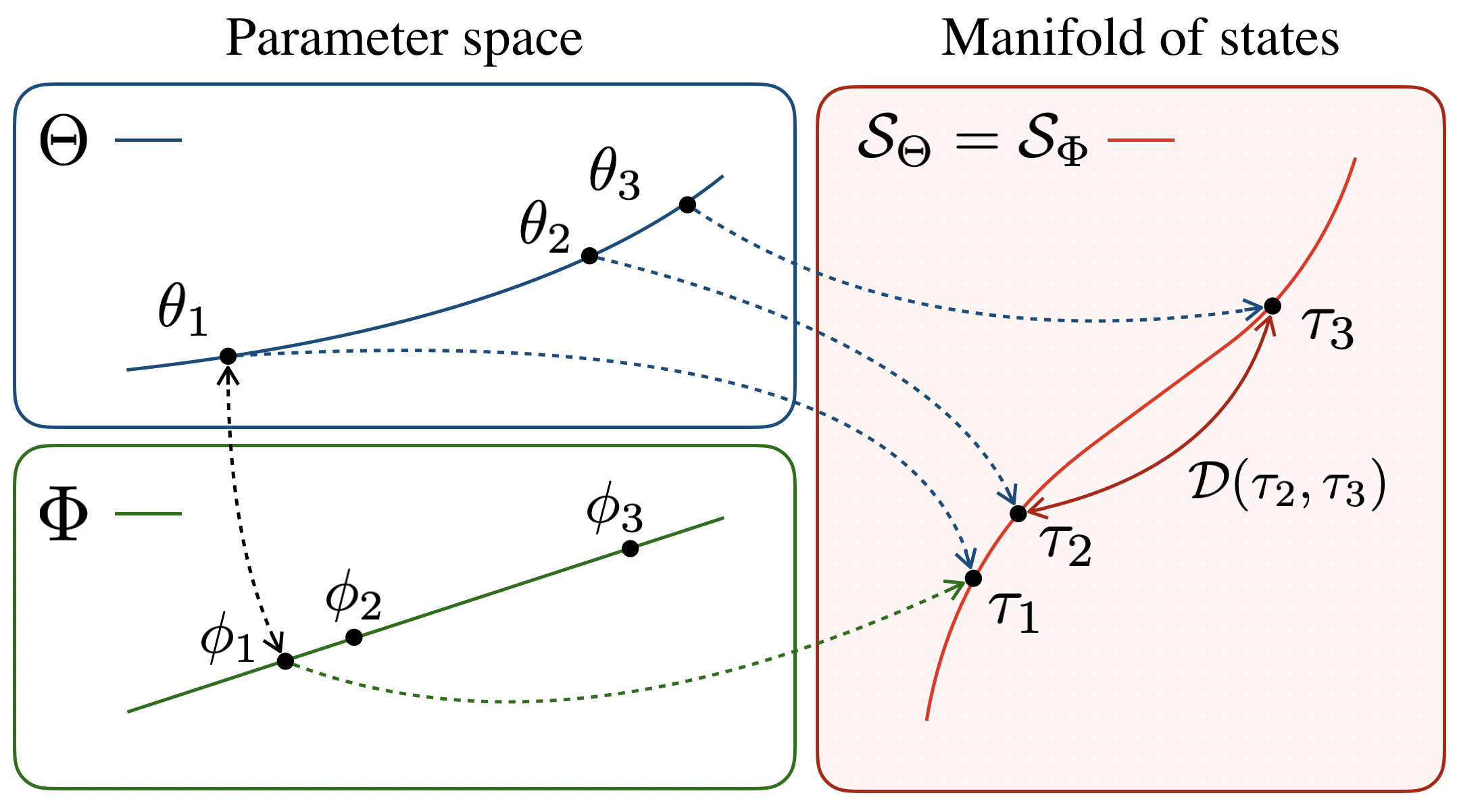}
    \caption{Illustration of a parameter space $\Theta$, or an alternative parameterization $\Phi$, mapped into a one-parameter curve of quantum states $\mathcal{S}_{\Theta}=\mathcal{S}_{\Phi}$. The distinguishability of parameter values is only meaningful, when expressed as a distance $\mathcal{D}$ between the corresponding quantum states,
    i.e.,~the state space induces a geometry on the parameter space.
    The induced geometry may be \textit{simpler} for certain choices of parameterization. In particular, there may exist a Euclidean parameterization with a flat geometry.
    By considering the distance directly between quantum states, it follows that the distance is invariant with respect to the specific choice of parameterization of the states.}
    \label{fig:illustration}
\end{figure}

Having introduced the manifold of quantum states, and discussed probability densities on the parameter space, we must face the subtlety of parameterization invariance.
In the above we work with a parameterization $\Theta$, however the manifold of quantum states itself is invariant with respect to the specific choice of parameterization.
For example, the manifold of thermal Gibbs states is the same whether we parameterize it using the temperature or the inverse temperature.
This fact is illustrated in Fig.~\ref{fig:illustration}.
In general we can express this invariance as follows: if we consider a one-to-one mapping $\phi\!:\!\Theta \!\rightarrow\! \Phi$, where $\Phi\subseteq \mathbb{R}$ is the image of the map,
then the function $\phi$ provides an equally valid parameterization of the manifold of states, i.e.,~$\mathcal{S}_\Phi\equiv\mathcal{S}_\Theta$ with
\begin{equation}
    \mathcal{S}_\Phi = \left\lbrace \tau_{\small{\Phi}}(\phi) \ \ \text{for} \ \ \phi\in \Phi \right\rbrace,
\end{equation}
and we explicitly indicate that each state is now given with respect to the parameterization $\Phi$, while the \emph{invariance condition} reads:
\begin{equation}
    \forall \phi:\,\phi = \phi(\theta),\quad \tau_{\Theta}(\theta) = \tau_{\small{\Phi}}(\phi),
\end{equation}
and expresses the parameterization invariance of every equivalent quantum state.
Furthermore, the probability assigned to a given region of state space must be independent of the specific parameterization employed.
Thus, to be consistent, the prior probability density must satisfy the invariance condition~\cite{Udo_2011,Jermyn_2005}:
\begin{equation} \label{eq:prior_invariance}
    \forall \phi:\,\phi = \phi(\theta),\quad d\theta p_{\Theta}(\theta) = d\phi p_{\Phi}(\phi)% \ \ \ \text{for} \ \ \phi = \phi(\theta)
\end{equation}
Since the likelihood function, $p(x|\theta)$, only depends on the state itself, it is inherently parameterization invariant.
Hence, it thus follows that if the above invariance condition holds for the prior density it will also hold for the posterior density.
To simplify our notation we will for the most part not indicate the parameterization explicitly in what follows.
The choice of parameterization is implicitly indicated by the parameter value itself.

\subsection{Metric structure and the geodesic length}
The essential ingredient required for the development of a parameterization-invariant estimation theory on a Riemannian manifold is the metric structure, i.e.,~an infinitesimal notion of length,
on the manifold of quantum states~\cite{Jermyn_2005,Amari2007,Amari_1985}.
A metric makes it possible to define the distance between states on the manifold,
and more fundamentally, the idea of defining a probability density on a continuous parameter space, is not well-defined in the absence of such a metric structure~\cite{Jermyn_2005}.
Suppose for a moment that the manifold possesses a metric denoted by~$g$.
Then, it is always possible to construct a well-defined distance function, as the geodesic length -- which we refer to as the \emph{metric-based distance} -- defined by~\cite{Jermyn_2005,Amari_1985,Crooks_2007_thermo_length}
\begin{equation}
    \mathcal{D}(\tau(\theta_{0}),\tau(\theta_{1})) \coloneqq \abs{ \int_{\theta_{0}}^{\theta_{1}} d\theta \ g_{\Theta}(\theta) } ,
    \label{eq:d_geo} 
\end{equation}
where the quantity $d\theta g_{\Theta}(\theta)$ provides a parameterization-invariant integration measure, i.e., $d\theta g_{\Theta}(\theta) = d\phi g_{\Phi}(\phi)$ for $\phi=\phi(\theta)$.
The metric-based distance is a statistical distance between quantum states,
and it is a parameterization-invariant quantity -- this is depicted in Fig.~\ref{fig:illustration}.
To simplify our notation, in what follows we will write the metric-based distance as $\mathcal{D}(\theta_{0},\theta_{1})$, referring only to the parameter values.
However, it should be kept in mind that this is a distance between quantum states.
The form of the geodesic length above is valid for one-parameter problems. More generally, defining the geodesic length involves a minimization over paths~\cite{Jermyn_2005}.

The above way of defining a distance function on the parameter space may still seem ambiguous unless a particular choice of
metric can be justified. However, if we consider a reference measurement of the sample with $\small{\text{POVM}}$ elements $\mathcal{M}(y)$ for $y\in Y$, then a key insight of Bayesian information geometry~\cite{SNOUSSI2007163,Amari_1985,Jermyn_2005,Crooks_2007_thermo_length} is that the likelihood associated with this measurement, induces a metric of the form $g^{2}=h_{\mathcal{M}}$,
where $h_{\mathcal{M}}$ is the Fisher information (FI) associated with the $\mathcal{M}$ measurement~\cite{Jermyn_2005,Amari_1985,Paris_2009}
\begin{equation}
    h_{\mathcal{M}}(\theta) = \int_{Y} dy \ p(y\!\mid\! \theta) \left( \partial_{\theta} \log p(y\!\mid\! \theta) \right)^{2} .
\end{equation}
From the point of view of Bayesian probability theory, the Fisher information metric with respect to the reference measurement, is the unique Riemannian metric which satisfies both parameterization-invariance and any other invariance property of the likelihood function~\cite{Jermyn_2005,Jarzyne_2020}. Moreover, according to Chentsov's theorem, any other (monotonic) metric on the parameter space corresponds to the Fisher information metric, with respect to some reference measurement, up to a multiplicative constant~\cite{Jarzyne_2020,Chentsov_1978}.

\subsubsection{Quantum Fisher information metric}
The choice of reference measurement represents a degree of freedom, i.e.,~we can define the metric-based distance, and the estimation theory itself, relative to an arbitrary reference measurement of the sample.
A natural choice would be to maximize the distance $\mathcal{D}(\theta_{0},\theta_{1})$ over all possible measurements,
however this procedure does not generally yield a unique reference measurement.
To see this, we consider the problem of maximizing the metric at a specific parameter value.
This local maximization problem can be solved analytically,
and yields a metric defined relative to a projective measurement of the symmetric logarithmic derivative $\mathcal{L}(\theta)$ associated with the manifold of states~\cite{Braunstein_1994}. 
The symmetric logarithmic derivative ($\text{\footnotesize{SLD}}$) is defined implicitly by the relation~\cite{Braunstein_1994}:
\begin{equation}\label{eq:SLD}
    \mathcal{L}(\theta)\tau(\theta) + \tau(\theta) \mathcal{L}(\theta) = 2 \partial_{\theta} \tau(\theta) ,
\end{equation}
and is in general a function of the parameter value.
The problem alluded to above is then that the $\text{\footnotesize{SLD}}$ defines a natural reference measurement only if the eigenbasis of the $\text{\footnotesize{SLD}}$ is parameter independent,
i.e.,~the projectors must be parameter independent. Notice that this need not be the case for the $\text{\footnotesize{SLD}}$ eigenvalues.

If we adopt the $\text{\footnotesize{SLD}}$ reference measurement, then with this choice of reference we obtain the \emph{quantum Fisher information} ($\small{\text{QFI}}$) metric,
which gives the maximal likelihood-induced metric-based distance between sample states.
The $\small{\text{QFI}}$ metric is given by~\cite{Braunstein_1994}:
\begin{equation}
    \sqrt{h_{\mathcal{L}}(\theta)} = \Tr[\mathcal{L}(\theta)^{2}\tau(\theta)]^{1/2}.
\end{equation}
The resulting metric is equal to four times the Bures metric~\cite{Paris_2009}, and is thus directly related to the so-called fidelity between infinitesimally separated states in the manifold.
If the family of states considered is a thermal state ensemble, then the metric-based distance is also called the thermodynamic length~\cite{Weinhold1975,Salamon1983,Crooks_2007_thermo_length,Scandi_2019_thermodynamiclength}.
In the remainder of this paper we will always consider the SLD reference measurement, written $\mathcal{L}$, and the $\small{\text{QFI}}$ metric following from that choice.

\subsection{Choosing the correct Bayesian error}
Having equipped the manifold with a metric structure, and outlined how to construct a metric-based distance function,
we can develop a parameterization-invariant estimation theory by considering the mean-square distance as a suitable Bayesian error-function~\cite{Jermyn_2005}.
Before proceeding with this development, we want to stress the basic logic of the approach.
For a specific thermal sample system, the QFI metric associated with the SLD reference measurement specifies the thermodynamic geometry,
i.e.~it specifies how to measure the distance between thermal sample states.
Only once the states are defined whose temperature should be distinguished (the manifold of states),
is it possible to define the correct corresponding Bayesian error function.

\subsubsection{Mean-square distance within the $\lambda_{\mathcal{L}}$ parameterization}
In this section we consider an implemented measurement $\Pi$ giving measurement data $x$, and the SLD reference measurement $\mathcal{L}$ defining the thermodynamic length.
Suppose we denote the estimator of the true parameter value constructed using the measurement data $x$ as $\tilde{\theta}(x)$.
Then, we may define the \emph{mean-square distance} based on \eqnref{eq:d_geo} as
\begin{equation}
    \text{\footnotesize{MSD}}(\tilde{\theta};x) \; \coloneqq \int_{\Theta} d\theta \ p(\theta\!\mid\! x) \ \mathcal{D}(\tilde\theta(x),\theta)^{2} ,
    \label{eq:msd}
\end{equation}
which provides a parameterization-invariant measure of the confidence assigned to the adopted parameter estimate.
The mean-square distance is defined with respect to the likelihood function $p(x|\theta)$ associated with the implemented measurement $\Pi$,
and it is defined with respect to the distance function associated with the SLD reference measurement $\mathcal{L}$.
Furthermore, note that the mean-square distance is conditional on the measurement data, i.e.,~it is a stochastic quantity.

For convenience we define, implicitly, the function $\lambda_{\mathcal{L}}(\theta)$ as the inverse derivative of the
QFI metric associated with the measurement $\mathcal{L}$ as~\cite{Jermyn_2005}
\begin{equation}\label{eq:definition_lambda}
    \partial_{\theta} \lambda_{\mathcal{L}}(\theta) \coloneqq h^{1/2}_{\mathcal{L}}(\theta) .
\end{equation}
Since the QFI is non-negative it follows that the function $\lambda_{\mathcal{L}}$ is monotonically increasing.
Furthermore, if we consider a change of parameterization $\theta\rightarrow\phi$,
and note that under such a transformation the QFI transforms as $h_{\mathcal{L}}(\theta) = [\partial_{\theta}\phi]^{2}h_{\mathcal{L}}(\phi)$,
then we see that it follows directly from the definition of $\lambda_{\mathcal{L}}$ that it is a parameterization-invariant quantity,
i.e.,~$\lambda_{\mathcal{L},\Theta}(\theta) = \lambda_{\mathcal{L},\Phi}(\phi)$ for $\phi=\phi(\theta)$.
Using the $\lambda_{\mathcal{L}}$ function it is always possible to %express %the distance function 
express the distance \eref{eq:d_geo} defined %on the manifold 
based on the QFI metric in the Euclidean form, i.e.:
\begin{gather}
    \mathcal{D}(\tilde\theta,\theta) = \abs{\lambda_{\mathcal{L}}(\tilde\theta) - \lambda_{\mathcal{L}}(\theta)} .
\end{gather}
This follows directly from an application of the fundamental theorem of calculus %since 
for $g_{\Theta}(\theta) = \sqrt{h_{\mathcal{L}}(\theta)}$.

If we constrain ourselves to reference measurements for which the QFI
is non-vanishing (except perhaps at isolated points, e.g.,~the boundaries of the parameter domain),
then $\lambda_{\mathcal{L}}$ itself constitutes a valid parameterization of the one-parameter family within the manifold.
Referring to the associated parameter space as $\Lambda_{\mathcal{L}}$, it follows that when working in this parameterization the $\text{\footnotesize{MSD}}$ takes the simple form:
\begin{equation}
    \text{\footnotesize{MSD}}(\tilde{\lambda};x) \; = \int_{\Lambda_{\mathcal{L}}} d\lambda \ p(\lambda \!\mid\! x) \ ( \tilde\lambda(x) - \lambda )^{2} ,
    \label{eq:msd_lambda}
\end{equation}
where $\tilde{\lambda}(x) = \lambda(\tilde{\theta}(x))$,
and we have made use of Eq.~\eqref{eq:prior_invariance}, i.e.,~the invariance property of the posterior probability density function.
In the above we dropped the subscript $\mathcal{L}$ when referring to the $\lambda$ parameter itself.
Furthermore, from here on we will not explicitly indicate the space which is integrated over, i.e.~$\Lambda_{\mathcal{L}}$. Note that the
$\Lambda_{\mathcal{L}}$-parameterization is special in that it is associated with a QFI equal to unity, i.e.,~a flat metric.
The form of the $\text{\footnotesize{MSD}}$ is then simply the standard Euclidean distance with respect to the parameterization $\Lambda_{\mathcal{L}}$.

The remaining open question is how to choose a suitable estimation function.
A natural choice, and the one primarily adopted here, is the mean of $\lambda$ over the posterior probability distribution
\begin{equation}
    \bar\lambda(x) \;\ \coloneqq \int d\lambda \ p(\lambda\!\mid\! x) \ \lambda  .
\end{equation}
It can be shown that this choice of estimation function is classically optimal, in the sense that it minimizes the mean-square distance (Eq.~\eref{eq:msd_lambda})~\cite{van2007bayesian,Jermyn_2005,Kay1993}.
For this reason the posterior mean of $\lambda$ is called the \textit{minimal mean-square distance} ($\text{\footnotesize{MMSD}}$) estimator~\cite{Jermyn_2005}.
Notice that since the $\text{\footnotesize{MSD}}$ is a parameterization-invariant quantity, we can then directly obtain the corresponding $\text{\footnotesize{MMSD}}$ estimate of
$\theta$ as $\bar{\theta}(x) = \lambda^{-1}\bar{\lambda}(x)$, where $\lambda^{-1}$ denotes the inverse of the function $\lambda$.

%Lastly, we mention that it is sometimes relevant to consider \jk{some} sub-optimal estimators. In particular, the parameterization-invariant form of the \jk{\emph{maximum-a-posteriori-probability} (MAP) estimator} is defined as the parameter value, denoted by $\theta_{\text{MAP}}$, for which the invariant probability density $\pi(\theta\!\mid\! x)/\sqrt{h_{\Lambda}}(\theta)$ is maximal~\cite{Jermyn_2005,Bassett_2019}.
%The $\text{\footnotesize{MAP}}$ estimator is a common choice, which typically converges to $\text{\footnotesize{MMSD}}$ estimator
%in the limit of a narrow prior\jk{~\cite{van2007bayesian}}.
%For any arbitrary choice of estimation function, the $\text{\footnotesize{MSD}}$ (Eq.~\eqref{eq:msd_lambda}) is lower bounded by the $\text{\footnotesize{MSD}}$
%with respect to the $\text{\footnotesize{MMSD}}$ estimator.

\subsubsection{The expected mean-square distance}
Having introduced the mean-square distance as a measure of confidence, the next problem is to identify a suitable criterion for optimal measurement design.
We immediately face the difficulty that the mean-square distance is a stochastic quantity, i.e.,~in identifying the optimal measurement one must consider the full ensemble
\begin{equation}
    \Xi \equiv \left\lbrace \text{\footnotesize{MSD}}(\bar{\lambda};x) \ \ \text{for} \ \ x\in X\right\rbrace .
\end{equation}
In an actual experiment we sample this ensemble according to the likelihood $p(x\!\mid\!\theta_{\text{true}})$ with respect to the true quantum state of the sample.
Since the true state is unknown a priori, we cannot gauge the expected confidence in an estimate resulting from a measurement based on the true likelihood.
Instead we must assume that the ensemble $\Xi$ is sampled according to the marginal density $p(x)$,
%\begin{equation}
%    p(x) = \int d\theta p(x\!\mid\! \theta) p(\theta) ,
%\end{equation}
giving the likelihood averaged over the prior density.

In the simplest case we suppose that the ensemble $\Xi$ is well captured by its average behaviour, which constitutes a deterministic quantity, typically referred to as the \textit{Bayesian mean-squared error}~\cite{Kay1993,van2007bayesian}. As in our work we interpret it as the average evaluated over the marginal density of the $\text{\footnotesize{MSD}}$~\eref{eq:msd_lambda},
we refer to it as the \emph{expected mean-square distance} ($\footnotesize{\text{EMSD}}$), i.e.:
\begin{align}
    \text{\footnotesize{EMSD}}[\bar{\lambda}] 
    &\coloneqq\int\!dx\; p(x)\; \text{\footnotesize{MSD}}(\tilde{\lambda};x) \\
    &= \int\! d\lambda\, dx \ p(\lambda , x) \ \left( \bar\lambda(x) - \lambda \right)^{2},
\end{align}
where we use square brackets to denote the choice of estimator. Hence, a natural criterion for determining the optimal measurement design is to minimise the $\text{\footnotesize{EMSD}}$. Note that the optimal measurement will generally be a functional of the prior probability distribution, or in other words the optimal choice of measurement depends on the prior knowledge.
%One can always define increasingly complex design criteria, e.g.,~we could search for the measurement with the minimal $\text{\footnotesize{EMSD}}$
%given a constraint on the fluctuations over the ensemble $\Xi$.
%For practical purposes it is often better to take a simple criterion which reduces the computational resources required in solving the associated optimization problem.
%This becomes especially true if measurement optimization is to be performed in real time.

\subsection{Connection to the local paradigm}
We now consider the local limit of a narrow posterior distribution.
In this limit the $\text{\footnotesize{EMSD}}$ is lower bounded, and typically well approximated, by the \textit{expected Cram\'{e}r-Rao bound} ($\footnotesize{\text{ECRB}}$)~\cite{Bacharach_2019_BCRB,Braunstein_1994}.
To derive this bound, we first rewrite the $\text{\footnotesize{EMSD}}$ as
\begin{equation} \label{eq:local_limit}
    \text{\footnotesize{EMSD}}[\tilde{\lambda}] = \int d\lambda p(\lambda) \Delta^{2}\tilde{\lambda} ,
\end{equation}
where we have decomposed the joint probability and defined the \textit{``frequentist'' mean-square error} ($\text{\footnotesize{MSE}}$)~\cite{Kay1993}
\begin{equation}
\begin{aligned}
    \Delta^{2}\tilde{\lambda} \coloneqq \int dx \ p(x|\lambda) [\tilde{\lambda}(x)-\lambda]^{2} . %\geqslant [h_{\Pi}(\lambda)]^{-1}
\end{aligned}
\end{equation}
Adopting an asymptotically unbiased estimator,
implies that the $\text{\footnotesize{MSE}}$ satisfies the Cram\'{e}r-Rao bound~\cite{Kay1993}, and upon substitution we obtain the $\footnotesize{\text{ECRB}}$:
\begin{equation}
\begin{aligned}
    \text{\footnotesize{EMSD}}[\tilde\lambda_{\text{ub}}] \geqslant \footnotesize{\text{ECRB}} 
    & \coloneqq \int d\lambda \frac{p(\lambda)}{h_{\Pi}(\lambda)} \\
    & \ = \int d\theta \ p(\theta) \frac{ h_{\mathcal{L}}(\theta)}{h_{\Pi}(\theta)},
\end{aligned}
\end{equation}
where $\tilde\lambda_{\text{ub}}$ denotes any unbiased estimator, and we recall that $h_{\Pi}$ is the Fisher information associated with the implemented measurement.
The final equality follows from a change of parameterization, and an application of \eqnref{eq:definition_lambda}.
The $\footnotesize{\text{ECRB}}$ is typically tight in the asymptotic limit, i.e.,~the limit of a large number of measurement repetitions~\cite{Bacharach_2019_BCRB}.

The expression for the $\footnotesize{\text{ECRB}}$ given here offers a generalization of the results reported in~\cite{Mok_2020,Rubio_2020_global} --- 
these corresponds to $h_{\mathcal{L}}(\theta) \propto 1$ and $h_{\mathcal{L}}(\theta) \propto \theta^{-2}$ respectively.
The $\footnotesize{\text{ECRB}}$ shows that the $\footnotesize{\text{EMSD}}$ converges to a generalized version of the relative error averaged over the prior probability,
i.e.,~relative to the metric structure of the manifold.
Note that the $\footnotesize{\text{ECRB}}$ shows that the noise-to-signal ratio is only meaningful if the geometry of the problem is scale invariant, i.e.~$h_{\mathcal{L}}(\theta) \propto \theta^{-2}$.
For general geometries a physically meaningful notion of relative error is given by the mean-square distance. This insight alters the data analysis even in the asymptotic limit.

\subsection{Bayesian Cram\'{e}r-Rao bounds}
As noted above, the expected thermometric performance is naturally quantified via the $\footnotesize{\text{EMSD}}$,
and the optimal measurement strategy is the one for which the $\text{\footnotesize{EMSD}}$ is minimal.
As an alternative to directly considering the $\footnotesize{\text{EMSD}}$, we can assess the expected performance by studying bounds on the attainable precision.
In many cases this is computationally advantageous.
The cost of considering lower bounds is that we do not know in general if the $\text{\footnotesize{EMSD}}$ saturates them.

An advantage of working in the $\Lambda_{\mathcal{M}}$-parameterization is that the $\footnotesize{\text{EMSD}}$ takes the form, as mentioned, of the $\footnotesize{\text{BMSE}}$.
The $\footnotesize{\text{EMSD}}$ can then be directly lower bounded by means of the so-called Van-Trees inequalities~\cite{van2007bayesian},
which are valid when $p(\lambda)$ and $\lambda p(\lambda)$ vanish at the boundaries of $\Lambda_{\mathcal{M}}$.
We first consider the tightened Bayesian Cram\'{e}r-Rao bound ($\text{\footnotesize{TBCRB}}$)~\cite{Bacharach_2019_BCRB,Yan_2018_entropy}
\begin{equation}
    \text{\footnotesize{EMSD}}[\bar\lambda] \geqslant \text{\footnotesize{TBCRB}} \coloneqq \int dx \ p(x) \mathcal{Q}(x)^{-1} ,
\end{equation}
given in terms of the Bayesian information
\begin{equation}
    \mathcal{Q}(x) = \int d\lambda \ p(\lambda \!\mid\! x) \left[ \partial_{\lambda} \log p(\lambda \!\mid\! x) \right]^{2} .
\end{equation}
The $\text{\footnotesize{EMSD}}$ saturates the $\text{\footnotesize{TBCRB}}$, i.e.,~the bound is tight,
when the posterior probability density takes the form of a Gaussian distribution in the $\Lambda_{\mathcal{M}}$-parameterization with inverse variance $\mathcal{Q}(x)$~\cite{Bacharach_2019_BCRB}.
According to the Laplace-Bernstein-von Mises theorem, this condition is satisfied in the asymptotic limit~\cite{Yan_2018_entropy}.

Evaluating the $\text{\footnotesize{TBCRB}}$ is in general no less demanding than computing the $\text{\footnotesize{EMSD}}$,
and our reason for starting with the $\text{\footnotesize{TBCRB}}$ is that it comes with a condition for when it is tight.
Often it is more convenient to consider the Bayesian Cram\'{e}r-Rao bound ($\text{\footnotesize{BCRB}}$),
which is obtained via an application of Jensen's inequality~\cite{Bacharach_2019_BCRB,Bacharach_2019_proceedinngs}
\begin{equation}
\begin{aligned}
    \text{\footnotesize{TBCRB}}^{-1} & \leqslant \text{\footnotesize{BCRB}}^{-1}  \\
    & \coloneqq Q_{\text{prior}} + \int d\theta \ p(\theta) \frac{h_{\Pi}(\theta)}{h_{\mathcal{M}}(\theta)} ,
\end{aligned}
\end{equation}
%where $h_{\Pi}$ is the Fisher information associated with the implemented measurement, $h_{\Lambda}$ is the Fisher information associated with the reference measurement,
where the prior information content is quantified by
\begin{equation}
    Q_{\text{prior}} = \int d\lambda \ p(\lambda) \ \left[ \partial_{\lambda} \log p(\lambda) \right]^{2} ,
\end{equation}
which is simply the Bayesian information evaluated for the initial prior.
The $\text{\footnotesize{TBCRB}}$ saturates the $\text{\footnotesize{BCRB}}$ if the Bayesian information function is a constant independent of the measurement data.
The advantage of working in the $\Lambda_{\mathcal{M}}$-parameterization is that the derived bound is automatically parameterization invariant.

Both the $\text{\footnotesize{TBCRB}}$ and the $\text{\footnotesize{BCRB}}$ are applicable at the single-shot level of the measurement protocol.
In many cases the $\text{\footnotesize{TBCRB}}$ is as difficult to compute as the $\text{\footnotesize{EMSD}}$,
this leaves $\text{\footnotesize{BCRB}}$-optimization as a viable choice of optimization strategy.

\begin{figure}
    \centering
    \includegraphics[width=0.475\textwidth]{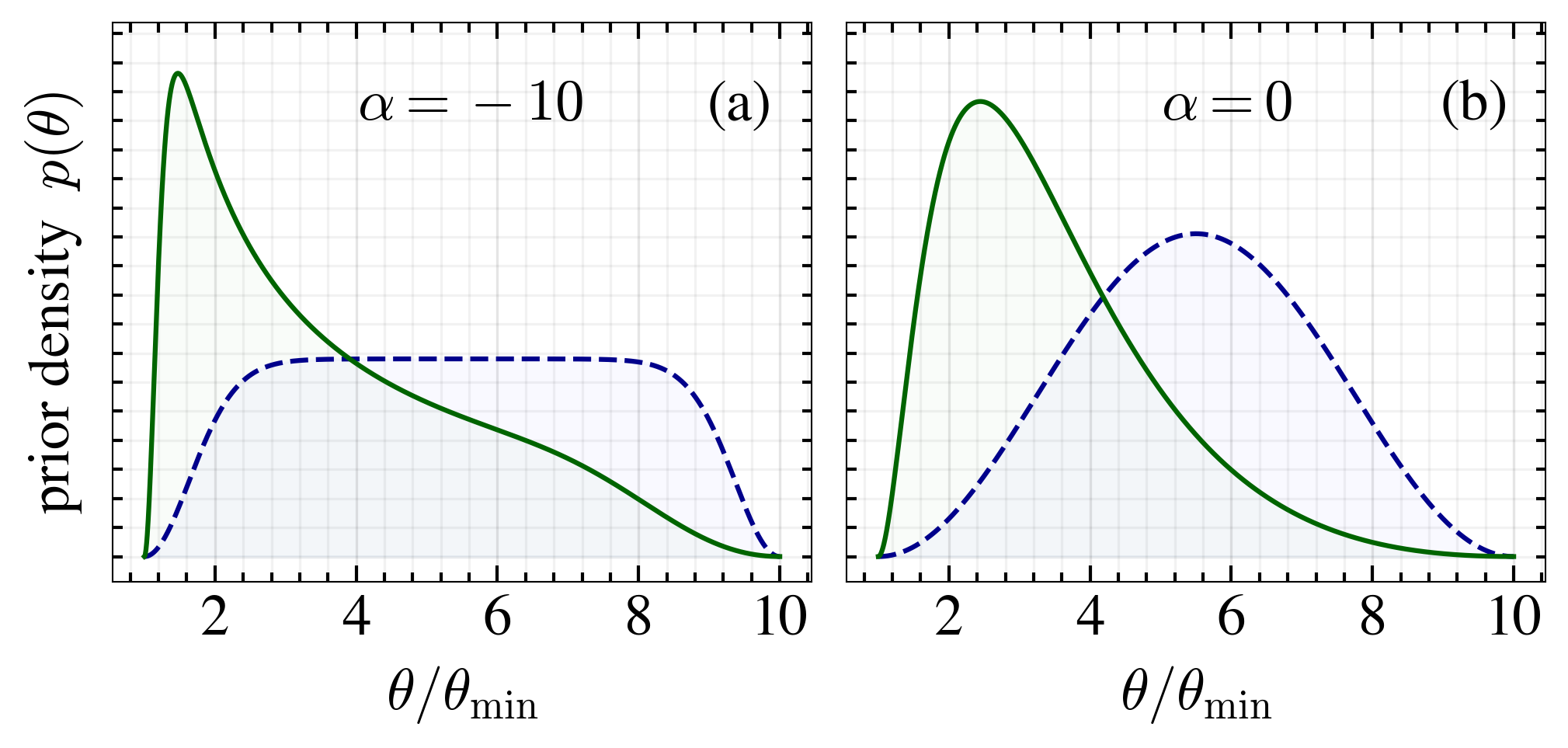}
    \caption{Illustration of the smoothed Jeffrey's prior (Eqs.~\eqref{eq:init_prior} and \eqref{eq:init_density})
    for a flat reference $h_{\mathcal{L}}(\theta)\propto 1$ (dashed blue lines) and a curved metric $h_{\mathcal{L}}(\theta)\propto 1/\theta^{2}$ (solid green lines)
    (a) Plot for $\alpha = -10$, in this limit we approach an equal a priori probability postulate on the bounded interval $\theta\in [\theta_{\min},\theta_{\max}]$.
    For a flat metric this gives a constant prior density with smoothed boundaries.
    For the curved metric the prior density is concentrated at small parameter values as the metric assigns the majority of metric length to this regime.
    (b) Plot for $\alpha=0$. In this regime, the prior resembles a Gaussian distribution on a bounded interval.}
    \label{fig:prior_illustration}
\end{figure}

\subsection{Specifying the initial prior}
Within the Bayesian approach, prior densities are updated to posterior densities as measurement data is acquired.
Applying the Bayesian framework requires specifying an initial prior, and it is crucial to choose a prior which correctly represents ones prior knowledge,
or does not strongly effect the inference process.
If we recall that the QFI metric encodes the thermodynamic geometry of the sample space, then
it is convenient to express a general prior as a product
\begin{equation}\label{eq:init_prior}
    p(\theta) = h_{\mathcal{L}}^{1/2}(\theta) f(\theta) ,
\end{equation}
where $f(\theta)$ is a parameterization-invariant density.
If we are interested in applying an equal a priori probability postulate on a specific domain of parameter space, i.e.,~we want an uninformative prior over the specified domain,
then this corresponds to having a constant, $f(\theta) = \text{cons.}$, prior density on this domain.
For a constant density the prior is proportional to the QFI metric, or in other words,
the probability assigned to an infinitesimal parameter interval is proportional to the length of the interval.
In this case we recover the so-called Jeffrey's prior, which is the prior expressing complete ignorance given the metric structure of the parameter space~\cite{Caticha_2008}.
For all simulations performed in the next section we adopt a smoothed density
\begin{equation}\label{eq:init_density}
    f(\theta) =  \frac{1}{\mathcal{N}}\left( \exp\left[\alpha \sin^{2}\left( \pi \frac{\lambda(\theta) - \lambda_{\min}}{\lambda_{\max}-\lambda_{\min}} \right)\right]-1 \right)
\end{equation}
with a normalization factor given by
\begin{equation}
    \mathcal{N} \equiv (\lambda_{\max}-\lambda_{\min}) \left[\exp(\alpha/2)I_{0}(\alpha/2)-1\right]
\end{equation}
where $\theta \in [\theta_{\min},\theta_{\max}]$, $\lambda_{\text{min,max}} = \lambda(\theta_{\text{min,max}})$ respectively and $I_{0}$ is the modified Bessel function of the first kind.
The resulting prior is essentially the one studied by Yan et al.~\cite{Yan_2018_entropy}, however we specify the density with respect to the Euclidean parameterization.
In the limit of large negative $\alpha$, the density goes to a constant on the parameter domain, and we thus recover Jeffrey's prior.
The prior is illustrated in figure~\ref{fig:prior_illustration} for a flat reference metric $h_{\mathcal{L}}(\theta)\propto 1$
and a curved reference metric given by $h_{\mathcal{L}}(\theta)\propto \theta^{-2}$.

%Briefly we note that in the case of a flat metric it follows that $\lambda(\theta) \propto \theta$,
%which implies that the mean-square distance with respect to $\mathcal{D}(\tilde\theta,\theta) = \mid\!\tilde\theta-\theta \!\mid$ is a well-defined quality measure.
%In phase estimation problems where the parameter is encoded via a unitary rotation of a given initial state,
%the quantum Fisher information becomes a constant independent of the parameter~\cite{Paris_2009}.
%This justifies the use of the mean-square error for unitary phase estimation problems.

\section{Applications}
In this section, we consider some applications of the above estimation theory to the problem of thermometry. First, we argue for a suitable form of the reference metric when the sample system can be represented as a thermalizing channel, and show that in this case the relative error is given by the standard noise-to-signal ratio. Second, we consider the case of thermal spin-$1/2$ particles, and illustrate the difference between a thermometric scenario in which the spin-$1/2$ particles are considered as a thermometer for an underlying heat reservoir, and a scenario in which the spin-$1/2$ particles themselves constitute the sample-system of interest.
These two scenarios correspond to scenario B and scenario A illustrated in Fig.~\ref{fig:conceptual_illustration}.

\subsection{Preliminaries}
We consider the case where the one-parameter family of sample-system states is the thermal Gibbs states with respect to a Hamiltonian operator $H$, and take $\theta$ to denote the temperature.
Recall that the Gibbs state takes the form~\cite{Mehboudi_2019_review,Pasquale_2018}
\begin{equation}
    \rho(\theta) = \frac{\exp\left( - H/k_{B}\theta \right)}{\Tr \left[ \exp\left( - H/k_{B}\theta \right) \right]} .
\end{equation}
As our reference measurement we take a projective measurement of the $\text{\small{SLD}}$ operator, which for Gibbs states takes the form~\cite{Mehboudi_2019_review,Pasquale_2018,Correa_2015}
\begin{equation}
    \mathcal{L}(\theta) = \frac{H - \Tr[\tau(\theta)H] }{2k_{B}\theta^{2}},
\end{equation}
where $k_{B}$ is the Boltzmann constant.
Measuring the $\text{\small{SLD}}$ thus corresponds to a projective measurement of the sample-system energy.
Noting that the Hamiltonian operator is temperature independent, and that this feature carries over to the eigenbasis of the $\text{\small{SLD}}$,
it follows that a projective measurement of the $\text{\small{SLD}}$ operator provides a valid reference measurement.
For a projective energy measurement the associated %$\text{\small{QFI}}$
FI, or equivalently the QFI,
is then directly related to the sample-system \textit{heat capacity}~\cite{Mehboudi_2019_review,Pasquale_2018,Correa_2015}, i.e.:
\begin{equation}
    h_{\mathcal{L}}(\theta)
    = \frac{\partial_{\theta} \Tr[\tau(\theta)H]}{k_{B}\theta^{2}}
    \eqqcolon \frac{C(\theta; H)}{k_{B}\theta^{2}}.
\end{equation}
The square-root of the $\text{\small{QFI}}$ then provides a metric on the manifold of thermal Gibbs states.
Recall that the associated metric-based distance is called the thermodynamic length~\cite{Weinhold1975,Salamon1983,Crooks_2007_thermo_length,Scandi_2019_thermodynamiclength}.
Below we evaluate the thermodynamic length in the case of a bosonic mode and a spin-$1/2$ particle.
First, however, we discuss how to represent a thermalizing channel.

\subsection{The thermalizing channel}
A common scenario in quantum thermometry, is that of a quantum probe subject to a thermalizing channel.
A thermalizing channel can in general be modelled as induced by a sample-system which is effectively an 
infinitely large heat reservoir. Here we model such an ideal heat reservoir by a heat capacity which, either approximately or by definition,
equals a constant value $k_{B}\mathcal{V}$ across the range of relevant temperatures, i.e.,~the sample energy is a linear function of the temperature.
Note that a constant heat capacity corresponds to the QFI metric $h_{\mathcal{L}}(\theta)=k_{B}\mathcal{V}/\theta^{2}$.
In this case we can evaluate the thermodynamic length analytically and find
\begin{equation}
    \mathcal{D}(\theta_{0},\theta_{1}) = \ \abs{\log(\theta_{1}/\theta_{0})} ,
\end{equation}
where for convenience we set $\mathcal{V}=1$.
The form of the $\text{\small{MSD}}$ resulting from this distance function is called the \emph{mean-square logarithmic error} $(\text{\small{MSLE}})$.
The $\text{\small{MSLE}}$ can be adopted whenever it can be assumed that the manifold of thermal states is generated via a weak coupling to an infinite heat reservoir.
In practice this assumption might break down at low temperatures, and more fundamentally there are cases where thermal behaviour cannot be linked to an infinite heat reservoir,
e.g.,~the case of subsystem thermalization described by the eigenstate thermalization hypothesis~\cite{Brenes_2020,Ashida_2018_thermalization}.

The $\text{\small{MSLE}}$ was recently proposed by Rubio et al.~\cite{Rubio_2020_global} as a suitable measure of confidence,
in the special case of a reference measurement for which the associated likelihood function satisfies the \textit{scale-invariance} property
\begin{equation}
    p(x\!\mid\! \theta) = \frac{g(x/\theta)}{\int dx \ g(x/\theta)} ,
\end{equation}
where $x$ denotes the outcome from a projective measurement of the sample energy and $g(x/\theta)$ is a function only of the dimensionless ratio $x/\theta$.
This scale-invariance property of the likelihood function is satisfied for sample-systems with a constant density of states, or equivalently, a constant heat capacity.

When considering the $\text{\small{MSLE}}$, it follows that the reference $\text{\small{QFI}}$ takes the form $h_{\mathcal{L}}(\theta) = \theta^{-2}$,
and we then find that the associated $\text{\small{ECRB}}$ for an unbiased estimator is given by
\begin{equation}
    \text{\footnotesize{ECRB}} = \int d\theta \ \frac{ p(\theta) }{\theta^{2}h_{\Pi}(\theta)} .
\end{equation}
The quantity $\theta^{2}h_{\Pi}(\theta)$ provides an upper bound on the signal-to-noise ratio within the frequentist estimation paradigm~\cite{Potts_2019_fundamentallimits,Jorgensen_2020_TightBound}.
This shows that when the sample can be modelled as an ideal heat reservoir, the standard notion of relative error is recovered in the local limit where $p(\theta)$ is sharply peaked.
However, our analysis also points out that the standard relative error is not suitable unless the sample-system has an approximately constant heat capacity across the range of relevant temperature.
This condition typically breaks down at sufficiently low temperatures.

\begin{figure}
    \centering
    \includegraphics[width=0.475\textwidth]{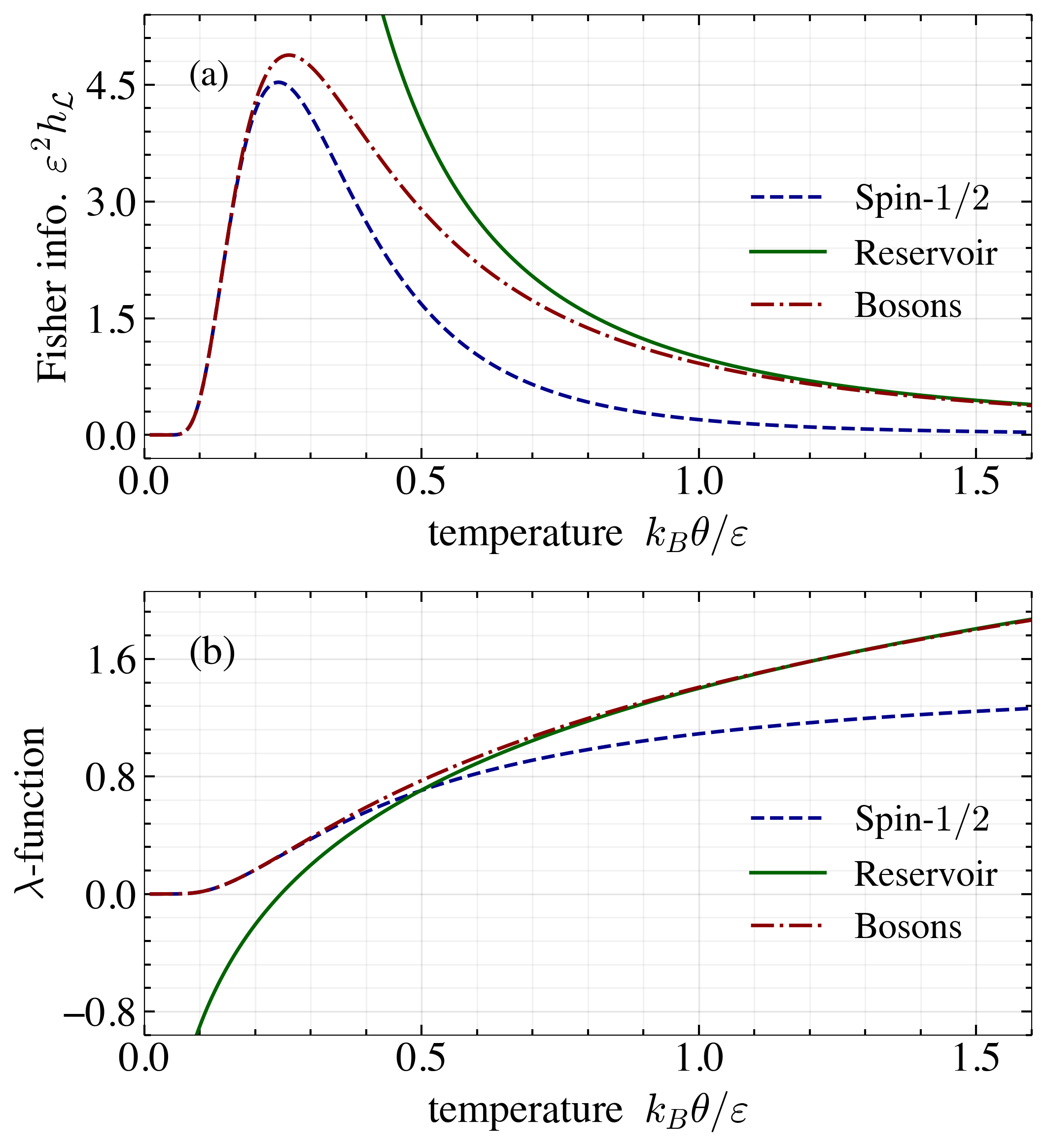}
    \caption{(a) Plot of the $\text{\small{QFI}}$ for a spin-$1/2$ particle (dashed blue line), the QFI for a bosonic mode (dashed-dotted red line),
    and the QFI for the ideal heat reservoir (solid green line).
    (b) Plot of the $\lambda$-functions associated with the various QFIs.}
    \label{fig:fisher_lambda}
\end{figure}

\subsubsection{Thermodynamic length for a Bosonic mode}
A specific system approximately realizing the above assumptions on an ideal heat reservoir is a Bose gas at a fixed density well above the critical temperature~\cite{Potts_2019_fundamentallimits}.
As an illustration we consider a gas of bosonic modes with energy gap $\varepsilon$. In this case the $\text{\small{QFI}}$ per mode is given by~\cite{Mehboudi_2019_review}
\begin{equation}
    h_{\mathcal{L},\text{boson}}(\theta) = \frac{\varepsilon^{2}/k_{B}^{2}\theta^{4}}{4 \sinh^{2}(\varepsilon/2k_{B}\theta)} .
\end{equation}
For this $\text{\small{QFI}}$ the associated $\lambda$-function (see \eqnref{eq:definition_lambda}) can be given an analytic expression
\begin{equation}
    \lambda_{\text{boson}}(\theta) = -\log\left[ \tanh \left( \varepsilon/4k_{B}\theta \right) \right] ,
\end{equation}
which implies that $\Lambda_{\mathcal{L},\text{boson}}=[0,+\infty)$.
The bosonic $\text{\small{QFI}}$ and the associated $\lambda$-function are shown in figure~\ref{fig:fisher_lambda}(a) and figure~\ref{fig:fisher_lambda}(b),
together with the corresponding quantities for an ideal heat reservoir.
We observe that in the limit where the temperature is large compared to the boson energy gap, the bosonic modes approximate the ideal heat reservoir,
i.e.~$h_{\mathcal{L},\text{boson}}(\theta)\rightarrow 1/\theta^{2}$.
This suggests that we can generically represent an ideal thermalizing channel physically by a collection of low-frequency bosonic modes.

\begin{figure*}
    \centering
    \includegraphics[width=0.95\textwidth]{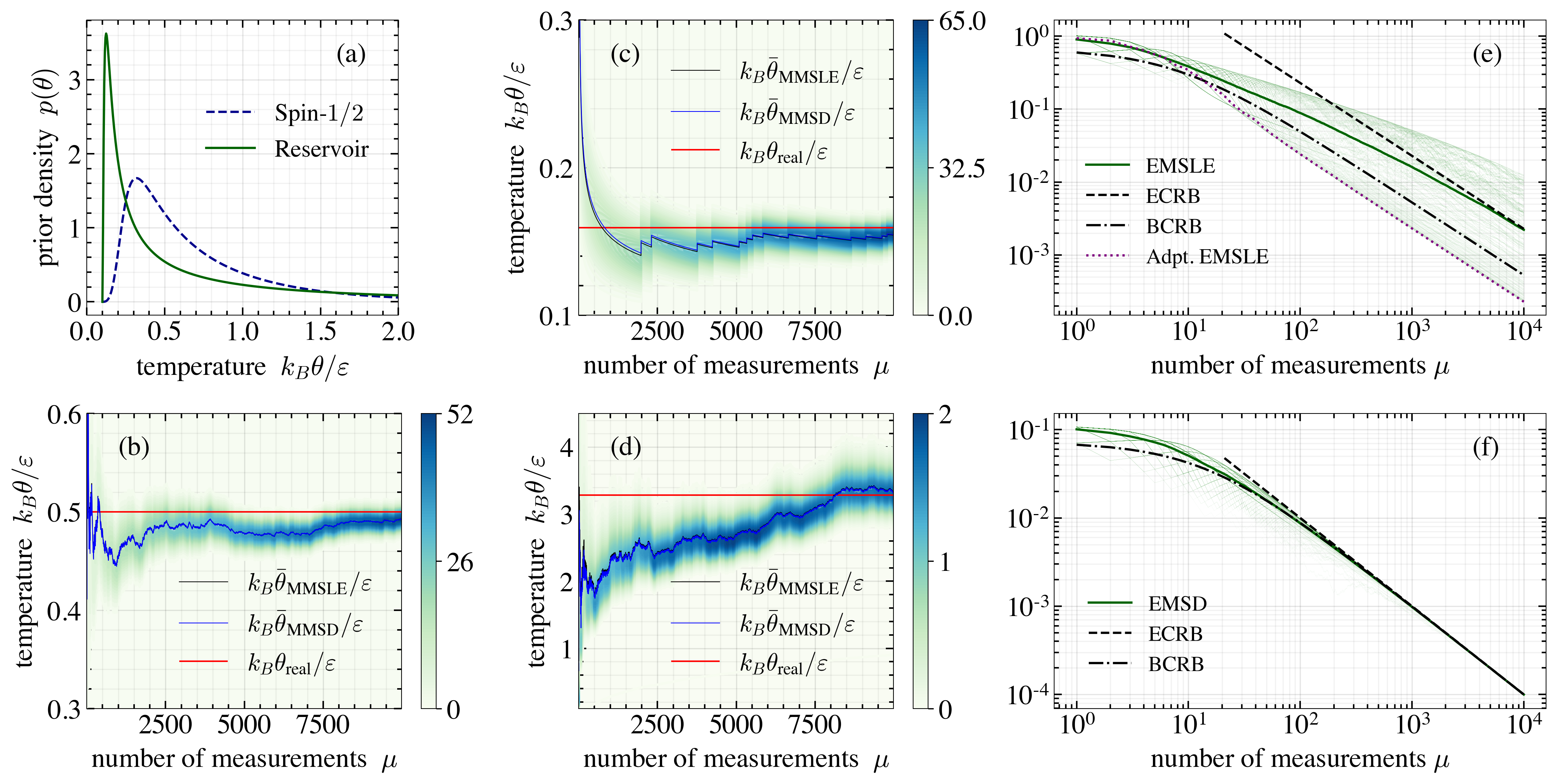}
    \caption{
    (a) The smoothed Jeffrey's prior (Eqs.~\eqref{eq:init_prior} and \eqref{eq:init_density}) for $\alpha = -4.5$ and $k_{B}\theta\in [\varepsilon/10,10\varepsilon]$,
    corresponding to the spin-$1/2$ particle and the reservoir reference metric.
    (b,c,d) Simulated stochastic measurement trajectories sampled according to a specific true temperature (red horizontal line),
    the posterior probability density function is shown as a colour map.
    Both the $\text{\footnotesize{MMSLE}}$ and the $\text{\footnotesize{MMSD}}$ estimators are shown.
    %(d) The smoothed Jeffrey's prior (Eqs.~\eqref{eq:init_prior} and \eqref{eq:init_density}) for $\alpha = -2.5$ and $k_{B}\theta\in [\varepsilon/10,5\varepsilon]$,
    %corresponding to the spin-$1/2$ particle and the reservoir reference metric.
    (e) Simulation of the $\text{\footnotesize{EMSLE}}$ (Scennario B) for thermometry employing free spin-$1/2$ particles.
    The simulation is performed by averaging over $250$ stochastic trajectories where the measurement data is generated with respect to a true temperature sampled according to the prior density.
    The $\text{\footnotesize{EMSLE}}$ is compared with the $\text{\footnotesize{ECRB}}$, convergence is observed for $\mu \gtrsim 10^{3}$,
    and we note that the generated trajectories do not converge to the mean asymptotically.
    Furthermore, we plot the $\text{\footnotesize{BCRB}}$ and observe that this is not a tight bound.
    Lastly, we plot the adaptive $\text{\footnotesize{EMSLE}}$. This is obtained by adapting the energy gap $\varepsilon$ in each repetition of the measurement.
    Here, we employ a $\text{\footnotesize{BCRB}}$ optimization strategy which is described in more detail in the main text.
    (f) Simulation of the $\text{\footnotesize{EMSD}}$ (Scenario A) for thermometry of free spin-$1/2$ particles.
    The simulation is performed by averaging over $250$ stochastic trajectories where the measurement data is generated with respect to a true temperature sampled according to the prior density.
    The $\text{\footnotesize{EMSD}}$ is compared with the $\text{\footnotesize{BCRB}}$ and convergence is observed for $\mu \gtrsim 10^{2}$,
    furthermore we observe that the generated trajectories converge to the average.}
    \label{fig:fermion_msd}
\end{figure*}

\subsection{Non-interacting spin-$1/2$ particles}
We now consider $N$ non-interacting spin-$1/2$ particles, or qubits, with identical energy gaps $\varepsilon$.
The spin-$1/2$ particles are in a thermal Gibbs state, and as above we take the $\theta$ parameterization to be the temperature.
We are going to compare and contrast two scenarios:
(A)~the spin-$1/2$ particles themselves constitute the sample-system of interest,
e.g.,~we are interested in thermometry of the spin degrees of freedom of the ultracold atoms themselves~\cite{Hartke_2020,Weld_2009_SpinGradient},
and (B)~the spin-$1/2$ particles are employed as equilibrium thermometers of an underlying heat reservoir,
e.g.,~the particles could model impurities within an ultracold gas, mapping motional information of gas atoms onto the quantum-spin state~\cite{Bouton_2020,Olf_2015_BEC}.

In case (B), as argued above, the $\text{\footnotesize{MSLE}}$ is the suitable measure of confidence,
and the $\text{\small{QFI}}$ metric takes the form $h_{\mathcal{L},\text{res.}}(\theta) = \theta^{-2}$.
In case (A) we take the $\text{\footnotesize{MSD}}$ resulting from adopting the $\text{\small{QFI}}$ metric of a thermal spin-$1/2$ particle as our reference, i.e.,~\cite{Potts_2019_fundamentallimits}
\begin{equation}
    h_{\mathcal{L},\text{spin}}(\theta) = \frac{\varepsilon^{2}/k_{B}^{2}\theta^{4}}{4 \cosh^{2}(\varepsilon/2k_{B}\theta)} ,
\end{equation}
which, we recall, corresponds to a projective energy measurement of a single spin-$1/2$ particle.
When referring to the $\text{\footnotesize{MSD}}$ below, we refer to the spin-$1/2$ particle $\text{\footnotesize{QFI}}$ metric.
For the spin-$1/2$ reference, the inverse derivative function $\lambda(\theta)$ can be obtained analytically, it takes the form
\begin{equation}\label{eq:lambda_fermions}
    \lambda_{\text{spin}}(\theta) = \pi - 2\arctan\left(e^{\varepsilon/2k_{B}\theta}\right),
\end{equation}
and thus $\Lambda_{\mathcal{L},\text{spin}} = [0,\pi/2]$.
In figure~\ref{fig:fisher_lambda}(a) we plot the spin-$1/2$ $\text{\small{QFI}}$ and compare it with the heat reservoir reference,
and in figure~\ref{fig:fisher_lambda}(b) we plot the associated $\lambda$-functions.
We observe that for temperature around $k_{B}\theta/\varepsilon \approx 0.3$ the two $\lambda$-functions exhibit a similar gradient,
however away from this temperature regime the specific thermometric scenario considered plays a role.

The estimation strategy employed consists of projectively measuring the energy of a subset of $\mu$ particles -- this is written as number of measurements in Fig.~\ref{fig:fermion_msd}.
For thermalized spin-$1/2$ particles, projective energy measurements maximize the associated $\text{\small{FI}}$ for all temperatures~\cite{Paris_2015},
and it follows that the $\text{\small{FI}}$ associated with this measurement can be expressed as
\begin{equation}
    h_{\Pi}(\theta) = \mu h_{\mathcal{L},\text{spin}}(\theta) .
\end{equation}
In figure~\ref{fig:fermion_msd}(a), figure~\ref{fig:fermion_msd}(b) and figure~\ref{fig:fermion_msd}(c) we show stochastic simulations of measurement trajectories
sampled according to three different true temperatures.
In all cases we plot the $\text{\footnotesize{MMSLE}}$ estimator and the $\text{\footnotesize{MMSD}}$ estimator,
and note that only negligible differences exist between these, despite different priors being used.
This feature is to be expected as the respective $\lambda$-functions rapidly become approximately constant across the posterior,
when this is the case the estimator is simply the maximum-a-posterior temperature, i.e., the temperature at which the posterior takes its maximum value, independently of the specific $\lambda$-function.

Although the various temperature estimates are only negligibly different, the confidence assigned to the estimates depend on the thermometric scenario.
In figure~\ref{fig:fermion_msd}(e) and \ref{fig:fermion_msd}(f) we show the $\text{\footnotesize{EMSLE}}$ and the $\text{\footnotesize{EMSD}}$ respectively as a function of the subset size $\mu$,
or equivalently the number of independent measurements,
and compare these with the associated $\text{\footnotesize{BCRB}}$ and the associated $\text{\footnotesize{ECRB}}$.
The $\text{\footnotesize{EMSLE}}$ and the $\text{\footnotesize{EMSD}}$ are evaluated using the corresponding smoothed Jeffrey's priors shown in figure~\ref{fig:fermion_msd}(d).
In the case of the $\text{\footnotesize{EMSLE}}$ we do not observe convergence to the $\text{\footnotesize{BCRB}}$.
This is to be expected since $h_{\mathcal{L},\text{res.}}/h_{\Pi}$ is generally not constant across the domain of the employed prior.
This feature also means that we do not observe a convergence to be $\text{\footnotesize{ECRB}}$ at the trajectory level, i.e., we observe fluctuations around the average.
For the $\text{\footnotesize{EMSD}}$ we observe rapid convergence to the $\text{\footnotesize{BCRB}}$, and trajectory level convergence to the $\text{\footnotesize{ECRB}}$.
This is due to the fact that $h_{\mathcal{L},\text{spin}}/h_{\Pi} = 1/\mu$ is a constant independent of the temperature.

When considering the spin-$1/2$ particles as thermometers of a underlying heat reservoir, it is sensible to consider adaptation of the energy gap $\varepsilon$ to optimize the thermometric performance\footnote{
See the accompanying paper~\cite{Mehboudi_accompany} focussing explicitly on the role of adaptivity in saturating fundamental precision bounds on Bayesian quantum thermometry via equilibrium quantum probes.}.
That is we consider a protocol consisting of thermalizing a spin-$1/2$ particle and projectively measuring the energy.
Based on the observed outcome, the energy gap of the particle employed in the following measurement is adapted.
In figure~\ref{fig:fermion_msd}(e) we also show the adaptive $\text{\footnotesize{EMSLE}}$, which is obtained via a $\text{\footnotesize{BCRB}}$ optimization strategy of the spin-$1/2$ energy gap $\varepsilon$ in each iteration of the protocol. By a $\text{\footnotesize{BCRB}}$ optimization strategy we refer to a minimization of the $\text{\footnotesize{BCRB}}$ with respect to the energy gap $\varepsilon$ in each iteration. In this case we observe a rapid convergence to the optimized local result, i.e., the adaptive $\text{\footnotesize{EMSLE}} \sim \max_{\theta} [ h_{\mathcal{L},\text{res.}}/h_{\Pi}]$. Furthermore, the adaptive $\text{\footnotesize{EMSLE}}$ converges to a lower bound on the set of $\text{\footnotesize{MSLE}}$ trajectories.
This is to be expected as the energy gap can be tuned to the optimal ratio of the energy gap to the temperature sampled from the prior.

\section{Conclusion}
In this paper, we have developed a general Bayesian approach to quantum thermometry based on the concept of thermodynamic length.
The basic idea is that a meaningful measure of the precision of a temperature estimate should be based on the ability to distinguish states at different temperatures,
i.e., colder from hotter, and should be independent of the particular parameterization of the states, e.g., temperature.
The theory allows us to meaningfully quantify the thermometric performance of a given measurement strategy,
and to design optimal measurements under conditions of prior temperature uncertainty.
Furthermore, the framework is applicable at the single-shot level, which allows us to consider adaptive measurement schemes.
Insisting on a framework which satisfies both parameterization invariance and any other symmetry of the sample system considered leads us to gauge the thermometric performance using a distance function defined relative to the quantum Fisher information metric on the manifold of thermal states of the sample, i.e.,~the thermodynamic length.
Formulating the problem in terms of the thermodynamic length, then leads to the realization that any meaningful notion of relative error must be given with respect to the specific sample system considered.
Furthermore, we demonstrate that the suitability of both Bayesian and local Cram\'{e}r-Rao bounds is contingent on the specific sample-system which is measured.

The generalization of the notion of relative error is particularly important in the low-temperature regime, where it shows that the standard signal-to-noise ratio is not a good indicator of estimation precision,
and provides an alternative point of view on the problems of low-temperature quantum thermometry~\cite{Jorgensen_2020_TightBound,Potts_2019_fundamentallimits}.
In addition the generalization is important at the microscale, where we are often not interested in thermometry of ideal heat baths,
but rather in characterizing the thermodynamics of operating quantum devices such as quantum simulators.

%%%%%%%%%%%%%%%%%%%%%%%%%%%%%%%%%%%%%%%%%%%%%%%%%%%%%%%%%%%%%%%%%%%%%
\section*{Acknowledgements}
MRJ and JBB acknowledge financial support from the Independent Research Fund Denmark.
MM and MPL acknowledge financial support from the Swiss National Science Foundations (NCCR SwissMAP and Ambizione grant PZ00P2-186067).
JK acknowledges financial support from the Foundation for Polish Science within the ``Quantum Optical Technologies'' project carried out within the International Research Agendas programme co-financed by the European Union under the European Regional Development Fund.

\bibliography{bibliography}

\begin{thebibliography}{54}%
\makeatletter
\providecommand \@ifxundefined [1]{%
 \@ifx{#1\undefined}
}%
\providecommand \@ifnum [1]{%
 \ifnum #1\expandafter \@firstoftwo
 \else \expandafter \@secondoftwo
 \fi
}%
\providecommand \@ifx [1]{%
 \ifx #1\expandafter \@firstoftwo
 \else \expandafter \@secondoftwo
 \fi
}%
\providecommand \natexlab [1]{#1}%
\providecommand \enquote  [1]{``#1''}%
\providecommand \bibnamefont  [1]{#1}%
\providecommand \bibfnamefont [1]{#1}%
\providecommand \citenamefont [1]{#1}%
\providecommand \href@noop [0]{\@secondoftwo}%
\providecommand \href [0]{\begingroup \@sanitize@url \@href}%
\providecommand \@href[1]{\@@startlink{#1}\@@href}%
\providecommand \@@href[1]{\endgroup#1\@@endlink}%
\providecommand \@sanitize@url [0]{\catcode `\\12\catcode `\$12\catcode
  `\&12\catcode `\#12\catcode `\^12\catcode `\_12\catcode `\%12\relax}%
\providecommand \@@startlink[1]{}%
\providecommand \@@endlink[0]{}%
\providecommand \url  [0]{\begingroup\@sanitize@url \@url }%
\providecommand \@url [1]{\endgroup\@href {#1}{\urlprefix }}%
\providecommand \urlprefix  [0]{URL }%
\providecommand \Eprint [0]{\href }%
\providecommand \doibase [0]{http://dx.doi.org/}%
\providecommand \selectlanguage [0]{\@gobble}%
\providecommand \bibinfo  [0]{\@secondoftwo}%
\providecommand \bibfield  [0]{\@secondoftwo}%
\providecommand \translation [1]{[#1]}%
\providecommand \BibitemOpen [0]{}%
\providecommand \bibitemStop [0]{}%
\providecommand \bibitemNoStop [0]{.\EOS\space}%
\providecommand \EOS [0]{\spacefactor3000\relax}%
\providecommand \BibitemShut  [1]{\csname bibitem#1\endcsname}%
\let\auto@bib@innerbib\@empty
%</preamble>
\bibitem [{\citenamefont {Kucsko}\ \emph {et~al.}(2013)\citenamefont {Kucsko},
  \citenamefont {Maurer}, \citenamefont {Yao}, \citenamefont {Kubo},
  \citenamefont {Noh}, \citenamefont {Lo}, \citenamefont {Park},\ and\
  \citenamefont {Lukin}}]{Kucsko_2013}%
  \BibitemOpen
  \bibfield  {author} {\bibinfo {author} {\bibfnamefont {G.}~\bibnamefont
  {Kucsko}}, \bibinfo {author} {\bibfnamefont {P.~C.}\ \bibnamefont {Maurer}},
  \bibinfo {author} {\bibfnamefont {N.~Y.}\ \bibnamefont {Yao}}, \bibinfo
  {author} {\bibfnamefont {M.}~\bibnamefont {Kubo}}, \bibinfo {author}
  {\bibfnamefont {H.~J.}\ \bibnamefont {Noh}}, \bibinfo {author} {\bibfnamefont
  {P.~K.}\ \bibnamefont {Lo}}, \bibinfo {author} {\bibfnamefont
  {H.}~\bibnamefont {Park}}, \ and\ \bibinfo {author} {\bibfnamefont {M.~D.}\
  \bibnamefont {Lukin}},\ }\bibfield  {title} {\enquote {\bibinfo {title}
  {Nanometre-scale thermometry in a living cell},}\ }\href {\doibase
  10.1038/nature12373} {\bibfield  {journal} {\bibinfo  {journal} {Nature}\
  }\textbf {\bibinfo {volume} {500}},\ \bibinfo {pages} {54--58} (\bibinfo
  {year} {2013})}\BibitemShut {NoStop}%
\bibitem [{\citenamefont {Fujiwara}\ \emph {et~al.}(2020)\citenamefont
  {Fujiwara}, \citenamefont {Sun}, \citenamefont {Dohms}, \citenamefont
  {Nishimura}, \citenamefont {Suto}, \citenamefont {Takezawa}, \citenamefont
  {Oshimi}, \citenamefont {Zhao}, \citenamefont {Sadzak}, \citenamefont
  {Umehara}, \citenamefont {Teki}, \citenamefont {Komatsu}, \citenamefont
  {Benson}, \citenamefont {Shikano},\ and\ \citenamefont
  {Kage-Nakadai}}]{Fujiwaraeaba_2020_inVivo}%
  \BibitemOpen
  \bibfield  {author} {\bibinfo {author} {\bibfnamefont {Masazumi}\
  \bibnamefont {Fujiwara}}, \bibinfo {author} {\bibfnamefont {Simo}\
  \bibnamefont {Sun}}, \bibinfo {author} {\bibfnamefont {Alexander}\
  \bibnamefont {Dohms}}, \bibinfo {author} {\bibfnamefont {Yushi}\ \bibnamefont
  {Nishimura}}, \bibinfo {author} {\bibfnamefont {Ken}\ \bibnamefont {Suto}},
  \bibinfo {author} {\bibfnamefont {Yuka}\ \bibnamefont {Takezawa}}, \bibinfo
  {author} {\bibfnamefont {Keisuke}\ \bibnamefont {Oshimi}}, \bibinfo {author}
  {\bibfnamefont {Li}~\bibnamefont {Zhao}}, \bibinfo {author} {\bibfnamefont
  {Nikola}\ \bibnamefont {Sadzak}}, \bibinfo {author} {\bibfnamefont {Yumi}\
  \bibnamefont {Umehara}}, \bibinfo {author} {\bibfnamefont {Yoshio}\
  \bibnamefont {Teki}}, \bibinfo {author} {\bibfnamefont {Naoki}\ \bibnamefont
  {Komatsu}}, \bibinfo {author} {\bibfnamefont {Oliver}\ \bibnamefont
  {Benson}}, \bibinfo {author} {\bibfnamefont {Yutaka}\ \bibnamefont
  {Shikano}}, \ and\ \bibinfo {author} {\bibfnamefont {Eriko}\ \bibnamefont
  {Kage-Nakadai}},\ }\bibfield  {title} {\enquote {\bibinfo {title} {Real-time
  nanodiamond thermometry probing in vivo thermogenic responses},}\ }\href
  {\doibase 10.1126/sciadv.aba9636} {\bibfield  {journal} {\bibinfo  {journal}
  {Science Advances}\ }\textbf {\bibinfo {volume} {6}} (\bibinfo {year}
  {2020}),\ 10.1126/sciadv.aba9636}\BibitemShut {NoStop}%
\bibitem [{\citenamefont {Moreva}\ \emph {et~al.}(2020)\citenamefont {Moreva},
  \citenamefont {Bernardi}, \citenamefont {Traina}, \citenamefont {Sosso},
  \citenamefont {Tchernij}, \citenamefont {Forneris}, \citenamefont {Picollo},
  \citenamefont {Brida}, \citenamefont {Pastuovi\ifmmode~\acute{c}\else
  \'{c}\fi{}}, \citenamefont {Degiovanni}, \citenamefont {Olivero},\ and\
  \citenamefont {Genovese}}]{Moreva_2020}%
  \BibitemOpen
  \bibfield  {author} {\bibinfo {author} {\bibfnamefont {E.}~\bibnamefont
  {Moreva}}, \bibinfo {author} {\bibfnamefont {E.}~\bibnamefont {Bernardi}},
  \bibinfo {author} {\bibfnamefont {P.}~\bibnamefont {Traina}}, \bibinfo
  {author} {\bibfnamefont {A.}~\bibnamefont {Sosso}}, \bibinfo {author}
  {\bibfnamefont {S.~Ditalia}\ \bibnamefont {Tchernij}}, \bibinfo {author}
  {\bibfnamefont {J.}~\bibnamefont {Forneris}}, \bibinfo {author}
  {\bibfnamefont {F.}~\bibnamefont {Picollo}}, \bibinfo {author} {\bibfnamefont
  {G.}~\bibnamefont {Brida}}, \bibinfo {author} {\bibfnamefont {\ifmmode
  \check{Z}\else~\v{Z}\fi{}.}\ \bibnamefont {Pastuovi\ifmmode~\acute{c}\else
  \'{c}\fi{}}}, \bibinfo {author} {\bibfnamefont {I.~P.}\ \bibnamefont
  {Degiovanni}}, \bibinfo {author} {\bibfnamefont {P.}~\bibnamefont {Olivero}},
  \ and\ \bibinfo {author} {\bibfnamefont {M.}~\bibnamefont {Genovese}},\
  }\bibfield  {title} {\enquote {\bibinfo {title} {Practical applications of
  quantum sensing: A simple method to enhance the sensitivity of
  nitrogen-vacancy-based temperature sensors},}\ }\href {\doibase
  10.1103/PhysRevApplied.13.054057} {\bibfield  {journal} {\bibinfo  {journal}
  {Phys. Rev. Applied}\ }\textbf {\bibinfo {volume} {13}},\ \bibinfo {pages}
  {054057} (\bibinfo {year} {2020})}\BibitemShut {NoStop}%
\bibitem [{\citenamefont {Carcy}\ \emph {et~al.}(2021)\citenamefont {Carcy},
  \citenamefont {Herc\'e}, \citenamefont {Tenart}, \citenamefont {Roscilde},\
  and\ \citenamefont {Cl\'ement}}]{Carcy_2021}%
  \BibitemOpen
  \bibfield  {author} {\bibinfo {author} {\bibfnamefont {C\'ecile}\
  \bibnamefont {Carcy}}, \bibinfo {author} {\bibfnamefont {Ga\'etan}\
  \bibnamefont {Herc\'e}}, \bibinfo {author} {\bibfnamefont {Antoine}\
  \bibnamefont {Tenart}}, \bibinfo {author} {\bibfnamefont {Tommaso}\
  \bibnamefont {Roscilde}}, \ and\ \bibinfo {author} {\bibfnamefont {David}\
  \bibnamefont {Cl\'ement}},\ }\bibfield  {title} {\enquote {\bibinfo {title}
  {Certifying the adiabatic preparation of ultracold lattice bosons in the
  vicinity of the mott transition},}\ }\href {\doibase
  10.1103/PhysRevLett.126.045301} {\bibfield  {journal} {\bibinfo  {journal}
  {Phys. Rev. Lett.}\ }\textbf {\bibinfo {volume} {126}},\ \bibinfo {pages}
  {045301} (\bibinfo {year} {2021})}\BibitemShut {NoStop}%
\bibitem [{\citenamefont {McKay}\ and\ \citenamefont
  {DeMarco}(2011)}]{McKay_2011}%
  \BibitemOpen
  \bibfield  {author} {\bibinfo {author} {\bibfnamefont {D~C}\ \bibnamefont
  {McKay}}\ and\ \bibinfo {author} {\bibfnamefont {B}~\bibnamefont {DeMarco}},\
  }\bibfield  {title} {\enquote {\bibinfo {title} {Cooling in strongly
  correlated optical lattices: prospects and challenges},}\ }\href {\doibase
  10.1088/0034-4885/74/5/054401} {\bibfield  {journal} {\bibinfo  {journal}
  {Reports on Progress in Physics}\ }\textbf {\bibinfo {volume} {74}},\
  \bibinfo {pages} {054401} (\bibinfo {year} {2011})}\BibitemShut {NoStop}%
\bibitem [{\citenamefont {Mitchison}\ \emph {et~al.}(2020)\citenamefont
  {Mitchison}, \citenamefont {Fogarty}, \citenamefont {Guarnieri},
  \citenamefont {Campbell}, \citenamefont {Busch},\ and\ \citenamefont
  {Goold}}]{Mitchison_2020_InSitu}%
  \BibitemOpen
  \bibfield  {author} {\bibinfo {author} {\bibfnamefont {Mark~T.}\ \bibnamefont
  {Mitchison}}, \bibinfo {author} {\bibfnamefont {Thom\'as}\ \bibnamefont
  {Fogarty}}, \bibinfo {author} {\bibfnamefont {Giacomo}\ \bibnamefont
  {Guarnieri}}, \bibinfo {author} {\bibfnamefont {Steve}\ \bibnamefont
  {Campbell}}, \bibinfo {author} {\bibfnamefont {Thomas}\ \bibnamefont
  {Busch}}, \ and\ \bibinfo {author} {\bibfnamefont {John}\ \bibnamefont
  {Goold}},\ }\bibfield  {title} {\enquote {\bibinfo {title} {In situ
  thermometry of a cold fermi gas via dephasing impurities},}\ }\href {\doibase
  10.1103/PhysRevLett.125.080402} {\bibfield  {journal} {\bibinfo  {journal}
  {Phys. Rev. Lett.}\ }\textbf {\bibinfo {volume} {125}},\ \bibinfo {pages}
  {080402} (\bibinfo {year} {2020})}\BibitemShut {NoStop}%
\bibitem [{\citenamefont {Mehboudi}\ \emph
  {et~al.}(2019{\natexlab{a}})\citenamefont {Mehboudi}, \citenamefont {Lampo},
  \citenamefont {Charalambous}, \citenamefont {Correa}, \citenamefont
  {Garc\'{\i}a-March},\ and\ \citenamefont {Lewenstein}}]{Mehboudi_2019}%
  \BibitemOpen
  \bibfield  {author} {\bibinfo {author} {\bibfnamefont {Mohammad}\
  \bibnamefont {Mehboudi}}, \bibinfo {author} {\bibfnamefont {Aniello}\
  \bibnamefont {Lampo}}, \bibinfo {author} {\bibfnamefont {Christos}\
  \bibnamefont {Charalambous}}, \bibinfo {author} {\bibfnamefont {Luis~A.}\
  \bibnamefont {Correa}}, \bibinfo {author} {\bibfnamefont {Miguel~\'Angel}\
  \bibnamefont {Garc\'{\i}a-March}}, \ and\ \bibinfo {author} {\bibfnamefont
  {Maciej}\ \bibnamefont {Lewenstein}},\ }\bibfield  {title} {\enquote
  {\bibinfo {title} {Using polarons for sub-nk quantum nondemolition
  thermometry in a bose-einstein condensate},}\ }\href {\doibase
  10.1103/PhysRevLett.122.030403} {\bibfield  {journal} {\bibinfo  {journal}
  {Phys. Rev. Lett.}\ }\textbf {\bibinfo {volume} {122}},\ \bibinfo {pages}
  {030403} (\bibinfo {year} {2019}{\natexlab{a}})}\BibitemShut {NoStop}%
\bibitem [{\citenamefont {Hartke}\ \emph {et~al.}(2020)\citenamefont {Hartke},
  \citenamefont {Oreg}, \citenamefont {Jia},\ and\ \citenamefont
  {Zwierlein}}]{Hartke_2020}%
  \BibitemOpen
  \bibfield  {author} {\bibinfo {author} {\bibfnamefont {Thomas}\ \bibnamefont
  {Hartke}}, \bibinfo {author} {\bibfnamefont {Botond}\ \bibnamefont {Oreg}},
  \bibinfo {author} {\bibfnamefont {Ningyuan}\ \bibnamefont {Jia}}, \ and\
  \bibinfo {author} {\bibfnamefont {Martin}\ \bibnamefont {Zwierlein}},\
  }\bibfield  {title} {\enquote {\bibinfo {title} {Doublon-hole correlations
  and fluctuation thermometry in a fermi-hubbard gas},}\ }\href {\doibase
  10.1103/PhysRevLett.125.113601} {\bibfield  {journal} {\bibinfo  {journal}
  {Phys. Rev. Lett.}\ }\textbf {\bibinfo {volume} {125}},\ \bibinfo {pages}
  {113601} (\bibinfo {year} {2020})}\BibitemShut {NoStop}%
\bibitem [{\citenamefont {Brantut}\ \emph {et~al.}(2013)\citenamefont
  {Brantut}, \citenamefont {Grenier}, \citenamefont {Meineke}, \citenamefont
  {Stadler}, \citenamefont {Krinner}, \citenamefont {Kollath}, \citenamefont
  {Esslinger},\ and\ \citenamefont {Georges}}]{Brantut_2013}%
  \BibitemOpen
  \bibfield  {author} {\bibinfo {author} {\bibfnamefont {Jean-Philippe}\
  \bibnamefont {Brantut}}, \bibinfo {author} {\bibfnamefont {Charles}\
  \bibnamefont {Grenier}}, \bibinfo {author} {\bibfnamefont {Jakob}\
  \bibnamefont {Meineke}}, \bibinfo {author} {\bibfnamefont {David}\
  \bibnamefont {Stadler}}, \bibinfo {author} {\bibfnamefont {Sebastian}\
  \bibnamefont {Krinner}}, \bibinfo {author} {\bibfnamefont {Corinna}\
  \bibnamefont {Kollath}}, \bibinfo {author} {\bibfnamefont {Tilman}\
  \bibnamefont {Esslinger}}, \ and\ \bibinfo {author} {\bibfnamefont {Antoine}\
  \bibnamefont {Georges}},\ }\bibfield  {title} {\enquote {\bibinfo {title} {A
  thermoelectric heat engine with ultracold atoms},}\ }\href {\doibase
  10.1126/science.1242308} {\bibfield  {journal} {\bibinfo  {journal}
  {Science}\ }\textbf {\bibinfo {volume} {342}},\ \bibinfo {pages} {713--715}
  (\bibinfo {year} {2013})}\BibitemShut {NoStop}%
\bibitem [{\citenamefont {Bouton}\ \emph {et~al.}(2020)\citenamefont {Bouton},
  \citenamefont {Nettersheim}, \citenamefont {Adam}, \citenamefont {Schmidt},
  \citenamefont {Mayer}, \citenamefont {Lausch}, \citenamefont {Tiemann},\ and\
  \citenamefont {Widera}}]{Bouton_2020}%
  \BibitemOpen
  \bibfield  {author} {\bibinfo {author} {\bibfnamefont {Quentin}\ \bibnamefont
  {Bouton}}, \bibinfo {author} {\bibfnamefont {Jens}\ \bibnamefont
  {Nettersheim}}, \bibinfo {author} {\bibfnamefont {Daniel}\ \bibnamefont
  {Adam}}, \bibinfo {author} {\bibfnamefont {Felix}\ \bibnamefont {Schmidt}},
  \bibinfo {author} {\bibfnamefont {Daniel}\ \bibnamefont {Mayer}}, \bibinfo
  {author} {\bibfnamefont {Tobias}\ \bibnamefont {Lausch}}, \bibinfo {author}
  {\bibfnamefont {Eberhard}\ \bibnamefont {Tiemann}}, \ and\ \bibinfo {author}
  {\bibfnamefont {Artur}\ \bibnamefont {Widera}},\ }\bibfield  {title}
  {\enquote {\bibinfo {title} {Single-atom quantum probes for ultracold gases
  boosted by nonequilibrium spin dynamics},}\ }\href {\doibase
  10.1103/PhysRevX.10.011018} {\bibfield  {journal} {\bibinfo  {journal} {Phys.
  Rev. X}\ }\textbf {\bibinfo {volume} {10}},\ \bibinfo {pages} {011018}
  (\bibinfo {year} {2020})}\BibitemShut {NoStop}%
\bibitem [{\citenamefont {Gasparinetti}\ \emph {et~al.}(2015)\citenamefont
  {Gasparinetti}, \citenamefont {Viisanen}, \citenamefont {Saira},
  \citenamefont {Faivre}, \citenamefont {Arzeo}, \citenamefont {Meschke},\ and\
  \citenamefont {Pekola}}]{Gasparinetti_2015}%
  \BibitemOpen
  \bibfield  {author} {\bibinfo {author} {\bibfnamefont {S.}~\bibnamefont
  {Gasparinetti}}, \bibinfo {author} {\bibfnamefont {K.~L.}\ \bibnamefont
  {Viisanen}}, \bibinfo {author} {\bibfnamefont {O.-P.}\ \bibnamefont {Saira}},
  \bibinfo {author} {\bibfnamefont {T.}~\bibnamefont {Faivre}}, \bibinfo
  {author} {\bibfnamefont {M.}~\bibnamefont {Arzeo}}, \bibinfo {author}
  {\bibfnamefont {M.}~\bibnamefont {Meschke}}, \ and\ \bibinfo {author}
  {\bibfnamefont {J.~P.}\ \bibnamefont {Pekola}},\ }\bibfield  {title}
  {\enquote {\bibinfo {title} {Fast electron thermometry for ultrasensitive
  calorimetric detection},}\ }\href {\doibase 10.1103/PhysRevApplied.3.014007}
  {\bibfield  {journal} {\bibinfo  {journal} {Phys. Rev. Applied}\ }\textbf
  {\bibinfo {volume} {3}},\ \bibinfo {pages} {014007} (\bibinfo {year}
  {2015})}\BibitemShut {NoStop}%
\bibitem [{\citenamefont {Mecklenburg}\ \emph {et~al.}(2015)\citenamefont
  {Mecklenburg}, \citenamefont {Hubbard}, \citenamefont {White}, \citenamefont
  {Dhall}, \citenamefont {Cronin}, \citenamefont {Aloni},\ and\ \citenamefont
  {Regan}}]{Mecklenburg_2015}%
  \BibitemOpen
  \bibfield  {author} {\bibinfo {author} {\bibfnamefont {Matthew}\ \bibnamefont
  {Mecklenburg}}, \bibinfo {author} {\bibfnamefont {William~A.}\ \bibnamefont
  {Hubbard}}, \bibinfo {author} {\bibfnamefont {E.~R.}\ \bibnamefont {White}},
  \bibinfo {author} {\bibfnamefont {Rohan}\ \bibnamefont {Dhall}}, \bibinfo
  {author} {\bibfnamefont {Stephen~B.}\ \bibnamefont {Cronin}}, \bibinfo
  {author} {\bibfnamefont {Shaul}\ \bibnamefont {Aloni}}, \ and\ \bibinfo
  {author} {\bibfnamefont {B.~C.}\ \bibnamefont {Regan}},\ }\bibfield  {title}
  {\enquote {\bibinfo {title} {Nanoscale temperature mapping in operating
  microelectronic devices},}\ }\href {\doibase 10.1126/science.aaa2433}
  {\bibfield  {journal} {\bibinfo  {journal} {Science}\ }\textbf {\bibinfo
  {volume} {347}},\ \bibinfo {pages} {629--632} (\bibinfo {year}
  {2015})}\BibitemShut {NoStop}%
\bibitem [{\citenamefont {Halbertal}\ \emph {et~al.}(2016)\citenamefont
  {Halbertal}, \citenamefont {Cuppens}, \citenamefont {Shalom}, \citenamefont
  {Embon}, \citenamefont {Shadmi}, \citenamefont {Anahory}, \citenamefont
  {Naren}, \citenamefont {Sarkar}, \citenamefont {Uri}, \citenamefont {Ronen},
  \citenamefont {Myasoedov}, \citenamefont {Levitov}, \citenamefont
  {Joselevich}, \citenamefont {Geim},\ and\ \citenamefont
  {Zeldov}}]{Halbertal_2016}%
  \BibitemOpen
  \bibfield  {author} {\bibinfo {author} {\bibfnamefont {D.}~\bibnamefont
  {Halbertal}}, \bibinfo {author} {\bibfnamefont {J.}~\bibnamefont {Cuppens}},
  \bibinfo {author} {\bibfnamefont {M.~Ben}\ \bibnamefont {Shalom}}, \bibinfo
  {author} {\bibfnamefont {L.}~\bibnamefont {Embon}}, \bibinfo {author}
  {\bibfnamefont {N.}~\bibnamefont {Shadmi}}, \bibinfo {author} {\bibfnamefont
  {Y.}~\bibnamefont {Anahory}}, \bibinfo {author} {\bibfnamefont {H.~R.}\
  \bibnamefont {Naren}}, \bibinfo {author} {\bibfnamefont {J.}~\bibnamefont
  {Sarkar}}, \bibinfo {author} {\bibfnamefont {A.}~\bibnamefont {Uri}},
  \bibinfo {author} {\bibfnamefont {Y.}~\bibnamefont {Ronen}}, \bibinfo
  {author} {\bibfnamefont {Y.}~\bibnamefont {Myasoedov}}, \bibinfo {author}
  {\bibfnamefont {L.~S.}\ \bibnamefont {Levitov}}, \bibinfo {author}
  {\bibfnamefont {E.}~\bibnamefont {Joselevich}}, \bibinfo {author}
  {\bibfnamefont {A.~K.}\ \bibnamefont {Geim}}, \ and\ \bibinfo {author}
  {\bibfnamefont {E.}~\bibnamefont {Zeldov}},\ }\bibfield  {title} {\enquote
  {\bibinfo {title} {Nanoscale thermal imaging of dissipation in quantum
  systems},}\ }\href {\doibase 10.1038/nature19843} {\bibfield  {journal}
  {\bibinfo  {journal} {Nature}\ }\textbf {\bibinfo {volume} {539}},\ \bibinfo
  {pages} {407--410} (\bibinfo {year} {2016})}\BibitemShut {NoStop}%
\bibitem [{\citenamefont {Giazotto}\ \emph {et~al.}(2006)\citenamefont
  {Giazotto}, \citenamefont {Heikkil\"a}, \citenamefont {Luukanen},
  \citenamefont {Savin},\ and\ \citenamefont {Pekola}}]{Giazotto_2006}%
  \BibitemOpen
  \bibfield  {author} {\bibinfo {author} {\bibfnamefont {Francesco}\
  \bibnamefont {Giazotto}}, \bibinfo {author} {\bibfnamefont {Tero~T.}\
  \bibnamefont {Heikkil\"a}}, \bibinfo {author} {\bibfnamefont {Arttu}\
  \bibnamefont {Luukanen}}, \bibinfo {author} {\bibfnamefont {Alexander~M.}\
  \bibnamefont {Savin}}, \ and\ \bibinfo {author} {\bibfnamefont {Jukka~P.}\
  \bibnamefont {Pekola}},\ }\bibfield  {title} {\enquote {\bibinfo {title}
  {Opportunities for mesoscopics in thermometry and refrigeration: Physics and
  applications},}\ }\href {\doibase 10.1103/RevModPhys.78.217} {\bibfield
  {journal} {\bibinfo  {journal} {Rev. Mod. Phys.}\ }\textbf {\bibinfo {volume}
  {78}},\ \bibinfo {pages} {217--274} (\bibinfo {year} {2006})}\BibitemShut
  {NoStop}%
\bibitem [{\citenamefont {Karimi}\ \emph {et~al.}(2020)\citenamefont {Karimi},
  \citenamefont {Brange}, \citenamefont {Samuelsson},\ and\ \citenamefont
  {Pekola}}]{Karimi_2020_resolution}%
  \BibitemOpen
  \bibfield  {author} {\bibinfo {author} {\bibfnamefont {Bayan}\ \bibnamefont
  {Karimi}}, \bibinfo {author} {\bibfnamefont {Fredrik}\ \bibnamefont
  {Brange}}, \bibinfo {author} {\bibfnamefont {Peter}\ \bibnamefont
  {Samuelsson}}, \ and\ \bibinfo {author} {\bibfnamefont {Jukka~P.}\
  \bibnamefont {Pekola}},\ }\bibfield  {title} {\enquote {\bibinfo {title}
  {Reaching the ultimate energy resolution of a quantum detector},}\ }\href
  {\doibase 10.1038/s41467-019-14247-2} {\bibfield  {journal} {\bibinfo
  {journal} {Nature Communications}\ }\textbf {\bibinfo {volume} {11}},\
  \bibinfo {pages} {367} (\bibinfo {year} {2020})}\BibitemShut {NoStop}%
\bibitem [{\citenamefont {Giovannetti}\ \emph {et~al.}(2011)\citenamefont
  {Giovannetti}, \citenamefont {Lloyd},\ and\ \citenamefont
  {Maccone}}]{Giovannetti_2011}%
  \BibitemOpen
  \bibfield  {author} {\bibinfo {author} {\bibfnamefont {Vittorio}\
  \bibnamefont {Giovannetti}}, \bibinfo {author} {\bibfnamefont {Seth}\
  \bibnamefont {Lloyd}}, \ and\ \bibinfo {author} {\bibfnamefont {Lorenzo}\
  \bibnamefont {Maccone}},\ }\bibfield  {title} {\enquote {\bibinfo {title}
  {Advances in quantum metrology},}\ }\href {\doibase 10.1038/nphoton.2011.35}
  {\bibfield  {journal} {\bibinfo  {journal} {Nature Photonics}\ }\textbf
  {\bibinfo {volume} {5}},\ \bibinfo {pages} {222--229} (\bibinfo {year}
  {2011})}\BibitemShut {NoStop}%
\bibitem [{\citenamefont {Giovannetti}\ \emph {et~al.}(2006)\citenamefont
  {Giovannetti}, \citenamefont {Lloyd},\ and\ \citenamefont
  {Maccone}}]{Giovannetti_2006}%
  \BibitemOpen
  \bibfield  {author} {\bibinfo {author} {\bibfnamefont {Vittorio}\
  \bibnamefont {Giovannetti}}, \bibinfo {author} {\bibfnamefont {Seth}\
  \bibnamefont {Lloyd}}, \ and\ \bibinfo {author} {\bibfnamefont {Lorenzo}\
  \bibnamefont {Maccone}},\ }\bibfield  {title} {\enquote {\bibinfo {title}
  {Quantum metrology},}\ }\href {\doibase 10.1103/PhysRevLett.96.010401}
  {\bibfield  {journal} {\bibinfo  {journal} {Phys. Rev. Lett.}\ }\textbf
  {\bibinfo {volume} {96}},\ \bibinfo {pages} {010401} (\bibinfo {year}
  {2006})}\BibitemShut {NoStop}%
\bibitem [{\citenamefont {Mehboudi}\ \emph
  {et~al.}(2019{\natexlab{b}})\citenamefont {Mehboudi}, \citenamefont
  {Sanpera},\ and\ \citenamefont {Correa}}]{Mehboudi_2019_review}%
  \BibitemOpen
  \bibfield  {author} {\bibinfo {author} {\bibfnamefont {Mohammad}\
  \bibnamefont {Mehboudi}}, \bibinfo {author} {\bibfnamefont {Anna}\
  \bibnamefont {Sanpera}}, \ and\ \bibinfo {author} {\bibfnamefont {Luis~A}\
  \bibnamefont {Correa}},\ }\bibfield  {title} {\enquote {\bibinfo {title}
  {Thermometry in the quantum regime: recent theoretical progress},}\ }\href
  {\doibase 10.1088/1751-8121/ab2828} {\bibfield  {journal} {\bibinfo
  {journal} {Journal of Physics A: Mathematical and Theoretical}\ }\textbf
  {\bibinfo {volume} {52}},\ \bibinfo {pages} {303001} (\bibinfo {year}
  {2019}{\natexlab{b}})}\BibitemShut {NoStop}%
\bibitem [{\citenamefont {{De Pasquale}}\ and\ \citenamefont
  {{Stace}}(2018)}]{Pasquale_2018}%
  \BibitemOpen
  \bibfield  {author} {\bibinfo {author} {\bibfnamefont {Antonella}\
  \bibnamefont {{De Pasquale}}}\ and\ \bibinfo {author} {\bibfnamefont
  {Thomas~M.}\ \bibnamefont {{Stace}}},\ }\enquote {\bibinfo {title} {{Quantum
  Thermometry}},}\ in\ \href {\doibase 10.1007/978-3-319-99046-0\_21} {\emph
  {\bibinfo {booktitle} {Thermodynamics in the Quantum Regime: Fundamental
  Aspects and New Directions}}},\ Vol.\ \bibinfo {volume} {195},\ \bibinfo
  {editor} {edited by\ \bibinfo {editor} {\bibfnamefont {Felix}\ \bibnamefont
  {{Binder}}}, \bibinfo {editor} {\bibfnamefont {Luis~A.}\ \bibnamefont
  {{Correa}}}, \bibinfo {editor} {\bibfnamefont {Christian}\ \bibnamefont
  {{Gogolin}}}, \bibinfo {editor} {\bibfnamefont {Janet}\ \bibnamefont
  {{Anders}}}, \ and\ \bibinfo {editor} {\bibfnamefont {Gerardo}\ \bibnamefont
  {{Adesso}}}}\ (\bibinfo {year} {2018})\ p.\ \bibinfo {pages}
  {503}\BibitemShut {NoStop}%
\bibitem [{\citenamefont {{Rubio}}\ \emph {et~al.}(2020)\citenamefont
  {{Rubio}}, \citenamefont {{Anders}},\ and\ \citenamefont
  {{Correa}}}]{Rubio_2020_global}%
  \BibitemOpen
  \bibfield  {author} {\bibinfo {author} {\bibfnamefont {Jes{\'u}s}\
  \bibnamefont {{Rubio}}}, \bibinfo {author} {\bibfnamefont {Janet}\
  \bibnamefont {{Anders}}}, \ and\ \bibinfo {author} {\bibfnamefont {Luis~A.}\
  \bibnamefont {{Correa}}},\ }\bibfield  {title} {\enquote {\bibinfo {title}
  {{Global Quantum Thermometry}},}\ }\href@noop {} {\bibfield  {journal}
  {\bibinfo  {journal} {arXiv e-prints}\ ,\ \bibinfo {eid} {arXiv:2011.13018}}
  (\bibinfo {year} {2020})},\ \Eprint {http://arxiv.org/abs/2011.13018}
  {arXiv:2011.13018 [quant-ph]} \BibitemShut {NoStop}%
\bibitem [{\citenamefont {{Alves}}\ and\ \citenamefont
  {{Landi}}(2021)}]{Alves_2021}%
  \BibitemOpen
  \bibfield  {author} {\bibinfo {author} {\bibfnamefont {Gabriel~O.}\
  \bibnamefont {{Alves}}}\ and\ \bibinfo {author} {\bibfnamefont {Gabriel~T.}\
  \bibnamefont {{Landi}}},\ }\bibfield  {title} {\enquote {\bibinfo {title}
  {{Bayesian estimation for collisional thermometry}},}\ }\href@noop {}
  {\bibfield  {journal} {\bibinfo  {journal} {arXiv e-prints}\ ,\ \bibinfo
  {eid} {arXiv:2106.12072}} (\bibinfo {year} {2021})},\ \Eprint
  {http://arxiv.org/abs/2106.12072} {arXiv:2106.12072 [quant-ph]} \BibitemShut
  {NoStop}%
\bibitem [{\citenamefont {{Mok}}\ \emph {et~al.}(2020)\citenamefont {{Mok}},
  \citenamefont {{Bharti}}, \citenamefont {{Kwek}},\ and\ \citenamefont
  {{Bayat}}}]{Mok_2020}%
  \BibitemOpen
  \bibfield  {author} {\bibinfo {author} {\bibfnamefont {Wai-Keong}\
  \bibnamefont {{Mok}}}, \bibinfo {author} {\bibfnamefont {Kishor}\
  \bibnamefont {{Bharti}}}, \bibinfo {author} {\bibfnamefont {Leong-Chuan}\
  \bibnamefont {{Kwek}}}, \ and\ \bibinfo {author} {\bibfnamefont {Abolfazl}\
  \bibnamefont {{Bayat}}},\ }\bibfield  {title} {\enquote {\bibinfo {title}
  {{Optimal Probes for Global Quantum Thermometry}},}\ }\href@noop {}
  {\bibfield  {journal} {\bibinfo  {journal} {arXiv e-prints}\ ,\ \bibinfo
  {eid} {arXiv:2010.14200}} (\bibinfo {year} {2020})},\ \Eprint
  {http://arxiv.org/abs/2010.14200} {arXiv:2010.14200 [quant-ph]} \BibitemShut
  {NoStop}%
\bibitem [{\citenamefont {Lehmann}\ and\ \citenamefont
  {Casella}(1998)}]{Lehmann1998}%
  \BibitemOpen
  \bibfield  {author} {\bibinfo {author} {\bibfnamefont {Erich~Leo}\
  \bibnamefont {Lehmann}}\ and\ \bibinfo {author} {\bibfnamefont {George}\
  \bibnamefont {Casella}},\ }\href@noop {} {\emph {\bibinfo {title} {Theory of
  Point Estimation}}},\ Vol.~\bibinfo {volume} {31}\ (\bibinfo  {publisher}
  {Springer},\ \bibinfo {year} {1998})\BibitemShut {NoStop}%
\bibitem [{\citenamefont {Kay}(1993)}]{Kay1993}%
  \BibitemOpen
  \bibfield  {author} {\bibinfo {author} {\bibfnamefont {Steven~M.}\
  \bibnamefont {Kay}},\ }\href@noop {} {\emph {\bibinfo {title} {Fundamentals
  of Statistical Signal Processing: Estimation Theory}}}\ (\bibinfo
  {publisher} {Prentice Hall},\ \bibinfo {year} {1993})\BibitemShut {NoStop}%
\bibitem [{\citenamefont {Paris}(2009)}]{Paris_2009}%
  \BibitemOpen
  \bibfield  {author} {\bibinfo {author} {\bibfnamefont {Matteo G.~A.}\
  \bibnamefont {Paris}},\ }\bibfield  {title} {\enquote {\bibinfo {title}
  {Quantum estimation for quantum technology},}\ }\href {\doibase
  10.1142/S0219749909004839} {\bibfield  {journal} {\bibinfo  {journal}
  {International Journal of Quantum Information}\ }\textbf {\bibinfo {volume}
  {07}},\ \bibinfo {pages} {125--137} (\bibinfo {year} {2009})},\ \Eprint
  {http://arxiv.org/abs/https://doi.org/10.1142/S0219749909004839}
  {https://doi.org/10.1142/S0219749909004839} \BibitemShut {NoStop}%
\bibitem [{\citenamefont {{Bacharach}}\ \emph {et~al.}(2019)\citenamefont
  {{Bacharach}}, \citenamefont {{Fritsche}}, \citenamefont {{Orguner}},\ and\
  \citenamefont {{Chaumette}}}]{Bacharach_2019_BCRB}%
  \BibitemOpen
  \bibfield  {author} {\bibinfo {author} {\bibfnamefont {Lucien}\ \bibnamefont
  {{Bacharach}}}, \bibinfo {author} {\bibfnamefont {Carsten}\ \bibnamefont
  {{Fritsche}}}, \bibinfo {author} {\bibfnamefont {Umut}\ \bibnamefont
  {{Orguner}}}, \ and\ \bibinfo {author} {\bibfnamefont {Eric}\ \bibnamefont
  {{Chaumette}}},\ }\bibfield  {title} {\enquote {\bibinfo {title} {{Some
  Results on Tighter Bayesian Lower Bounds on the Mean-Square Error}},}\
  }\href@noop {} {\bibfield  {journal} {\bibinfo  {journal} {arXiv e-prints}\
  ,\ \bibinfo {eid} {arXiv:1907.09509}} (\bibinfo {year} {2019})},\ \Eprint
  {http://arxiv.org/abs/1907.09509} {arXiv:1907.09509 [cs.IT]} \BibitemShut
  {NoStop}%
\bibitem [{\citenamefont {Braunstein}\ and\ \citenamefont
  {Caves}(1994)}]{Braunstein_1994}%
  \BibitemOpen
  \bibfield  {author} {\bibinfo {author} {\bibfnamefont {Samuel~L.}\
  \bibnamefont {Braunstein}}\ and\ \bibinfo {author} {\bibfnamefont
  {Carlton~M.}\ \bibnamefont {Caves}},\ }\bibfield  {title} {\enquote {\bibinfo
  {title} {Statistical distance and the geometry of quantum states},}\ }\href
  {\doibase 10.1103/PhysRevLett.72.3439} {\bibfield  {journal} {\bibinfo
  {journal} {Phys. Rev. Lett.}\ }\textbf {\bibinfo {volume} {72}},\ \bibinfo
  {pages} {3439--3443} (\bibinfo {year} {1994})}\BibitemShut {NoStop}%
\bibitem [{\citenamefont {von Toussaint}(2011)}]{Udo_2011}%
  \BibitemOpen
  \bibfield  {author} {\bibinfo {author} {\bibfnamefont {Udo}\ \bibnamefont
  {von Toussaint}},\ }\bibfield  {title} {\enquote {\bibinfo {title} {Bayesian
  inference in physics},}\ }\href {\doibase 10.1103/RevModPhys.83.943}
  {\bibfield  {journal} {\bibinfo  {journal} {Rev. Mod. Phys.}\ }\textbf
  {\bibinfo {volume} {83}},\ \bibinfo {pages} {943--999} (\bibinfo {year}
  {2011})}\BibitemShut {NoStop}%
\bibitem [{\citenamefont {Van~Trees}\ and\ \citenamefont
  {Bell}(2007)}]{van2007bayesian}%
  \BibitemOpen
  \bibfield  {author} {\bibinfo {author} {\bibfnamefont {H.~L.}\ \bibnamefont
  {Van~Trees}}\ and\ \bibinfo {author} {\bibfnamefont {K.~L.}\ \bibnamefont
  {Bell}},\ }\href@noop {} {\emph {\bibinfo {title} {{Bayesian Bounds for
  Parameter Estimation and Nonlinear Filtering/Tracking}}}}\ (\bibinfo
  {publisher} {Wiley},\ \bibinfo {year} {2007})\BibitemShut {NoStop}%
\bibitem [{\citenamefont {Weinhold}(1975)}]{Weinhold1975}%
  \BibitemOpen
  \bibfield  {author} {\bibinfo {author} {\bibfnamefont {F.}~\bibnamefont
  {Weinhold}},\ }\bibfield  {title} {\enquote {\bibinfo {title} {Metric
  geometry of equilibrium thermodynamics},}\ }\href {\doibase 10.1063/1.431689}
  {\bibfield  {journal} {\bibinfo  {journal} {The Journal of Chemical Physics}\
  }\textbf {\bibinfo {volume} {63}},\ \bibinfo {pages} {2479--2483} (\bibinfo
  {year} {1975})}\BibitemShut {NoStop}%
\bibitem [{\citenamefont {Salamon}\ and\ \citenamefont
  {Berry}(1983)}]{Salamon1983}%
  \BibitemOpen
  \bibfield  {author} {\bibinfo {author} {\bibfnamefont {Peter}\ \bibnamefont
  {Salamon}}\ and\ \bibinfo {author} {\bibfnamefont {R.~Stephen}\ \bibnamefont
  {Berry}},\ }\bibfield  {title} {\enquote {\bibinfo {title} {Thermodynamic
  length and dissipated availability},}\ }\href {\doibase
  10.1103/physrevlett.51.1127} {\bibfield  {journal} {\bibinfo  {journal}
  {Physical Review Letters}\ }\textbf {\bibinfo {volume} {51}},\ \bibinfo
  {pages} {1127--1130} (\bibinfo {year} {1983})}\BibitemShut {NoStop}%
\bibitem [{\citenamefont {Crooks}(2007)}]{Crooks_2007_thermo_length}%
  \BibitemOpen
  \bibfield  {author} {\bibinfo {author} {\bibfnamefont {Gavin~E.}\
  \bibnamefont {Crooks}},\ }\bibfield  {title} {\enquote {\bibinfo {title}
  {Measuring thermodynamic length},}\ }\href {\doibase
  10.1103/PhysRevLett.99.100602} {\bibfield  {journal} {\bibinfo  {journal}
  {Phys. Rev. Lett.}\ }\textbf {\bibinfo {volume} {99}},\ \bibinfo {pages}
  {100602} (\bibinfo {year} {2007})}\BibitemShut {NoStop}%
\bibitem [{\citenamefont {Scandi}\ and\ \citenamefont
  {Perarnau-Llobet}(2019)}]{Scandi_2019_thermodynamiclength}%
  \BibitemOpen
  \bibfield  {author} {\bibinfo {author} {\bibfnamefont {Matteo}\ \bibnamefont
  {Scandi}}\ and\ \bibinfo {author} {\bibfnamefont {Mart{\'{i}}}\ \bibnamefont
  {Perarnau-Llobet}},\ }\bibfield  {title} {\enquote {\bibinfo {title}
  {Thermodynamic length in open quantum systems},}\ }\href {\doibase
  10.22331/q-2019-10-24-197} {\bibfield  {journal} {\bibinfo  {journal}
  {{Quantum}}\ }\textbf {\bibinfo {volume} {3}},\ \bibinfo {pages} {197}
  (\bibinfo {year} {2019})}\BibitemShut {NoStop}%
\bibitem [{\citenamefont {Weld}\ \emph {et~al.}(2009)\citenamefont {Weld},
  \citenamefont {Medley}, \citenamefont {Miyake}, \citenamefont {Hucul},
  \citenamefont {Pritchard},\ and\ \citenamefont
  {Ketterle}}]{Weld_2009_SpinGradient}%
  \BibitemOpen
  \bibfield  {author} {\bibinfo {author} {\bibfnamefont {David~M.}\
  \bibnamefont {Weld}}, \bibinfo {author} {\bibfnamefont {Patrick}\
  \bibnamefont {Medley}}, \bibinfo {author} {\bibfnamefont {Hirokazu}\
  \bibnamefont {Miyake}}, \bibinfo {author} {\bibfnamefont {David}\
  \bibnamefont {Hucul}}, \bibinfo {author} {\bibfnamefont {David~E.}\
  \bibnamefont {Pritchard}}, \ and\ \bibinfo {author} {\bibfnamefont
  {Wolfgang}\ \bibnamefont {Ketterle}},\ }\bibfield  {title} {\enquote
  {\bibinfo {title} {Spin gradient thermometry for ultracold atoms in optical
  lattices},}\ }\href {\doibase 10.1103/PhysRevLett.103.245301} {\bibfield
  {journal} {\bibinfo  {journal} {Phys. Rev. Lett.}\ }\textbf {\bibinfo
  {volume} {103}},\ \bibinfo {pages} {245301} (\bibinfo {year}
  {2009})}\BibitemShut {NoStop}%
\bibitem [{\citenamefont {Olf}\ \emph {et~al.}(2015)\citenamefont {Olf},
  \citenamefont {Fang}, \citenamefont {Marti}, \citenamefont {MacRae},\ and\
  \citenamefont {Stamper-Kurn}}]{Olf_2015_BEC}%
  \BibitemOpen
  \bibfield  {author} {\bibinfo {author} {\bibfnamefont {Ryan}\ \bibnamefont
  {Olf}}, \bibinfo {author} {\bibfnamefont {Fang}\ \bibnamefont {Fang}},
  \bibinfo {author} {\bibfnamefont {G.~Edward}\ \bibnamefont {Marti}}, \bibinfo
  {author} {\bibfnamefont {Andrew}\ \bibnamefont {MacRae}}, \ and\ \bibinfo
  {author} {\bibfnamefont {Dan~M.}\ \bibnamefont {Stamper-Kurn}},\ }\bibfield
  {title} {\enquote {\bibinfo {title} {Thermometry and cooling of a bose gas to
  0.02 times the condensation temperature},}\ }\href {\doibase
  10.1038/nphys3408} {\bibfield  {journal} {\bibinfo  {journal} {Nature
  Physics}\ }\textbf {\bibinfo {volume} {11}},\ \bibinfo {pages} {720--723}
  (\bibinfo {year} {2015})}\BibitemShut {NoStop}%
\bibitem [{\citenamefont {Gallavotti}(1999)}]{Gallavotti_1999}%
  \BibitemOpen
  \bibfield  {author} {\bibinfo {author} {\bibfnamefont {Giovanni}\
  \bibnamefont {Gallavotti}},\ }\href@noop {} {\emph {\bibinfo {title}
  {Statistical Mechanics : A Short Treatise}}}\ (\bibinfo  {publisher}
  {Springer Berlin Heidelberg},\ \bibinfo {year} {1999})\BibitemShut {NoStop}%
\bibitem [{\citenamefont {Guarnieri}\ \emph {et~al.}(2019)\citenamefont
  {Guarnieri}, \citenamefont {Landi}, \citenamefont {Clark},\ and\
  \citenamefont {Goold}}]{Guarnieri_2019}%
  \BibitemOpen
  \bibfield  {author} {\bibinfo {author} {\bibfnamefont {Giacomo}\ \bibnamefont
  {Guarnieri}}, \bibinfo {author} {\bibfnamefont {Gabriel~T.}\ \bibnamefont
  {Landi}}, \bibinfo {author} {\bibfnamefont {Stephen~R.}\ \bibnamefont
  {Clark}}, \ and\ \bibinfo {author} {\bibfnamefont {John}\ \bibnamefont
  {Goold}},\ }\bibfield  {title} {\enquote {\bibinfo {title} {Thermodynamics of
  precision in quantum nonequilibrium steady states},}\ }\href {\doibase
  10.1103/PhysRevResearch.1.033021} {\bibfield  {journal} {\bibinfo  {journal}
  {Phys. Rev. Research}\ }\textbf {\bibinfo {volume} {1}},\ \bibinfo {pages}
  {033021} (\bibinfo {year} {2019})}\BibitemShut {NoStop}%
\bibitem [{\citenamefont {Brenes}\ \emph {et~al.}(2020)\citenamefont {Brenes},
  \citenamefont {Pappalardi}, \citenamefont {Goold},\ and\ \citenamefont
  {Silva}}]{Brenes_2020}%
  \BibitemOpen
  \bibfield  {author} {\bibinfo {author} {\bibfnamefont {Marlon}\ \bibnamefont
  {Brenes}}, \bibinfo {author} {\bibfnamefont {Silvia}\ \bibnamefont
  {Pappalardi}}, \bibinfo {author} {\bibfnamefont {John}\ \bibnamefont
  {Goold}}, \ and\ \bibinfo {author} {\bibfnamefont {Alessandro}\ \bibnamefont
  {Silva}},\ }\bibfield  {title} {\enquote {\bibinfo {title} {Multipartite
  entanglement structure in the eigenstate thermalization hypothesis},}\ }\href
  {\doibase 10.1103/PhysRevLett.124.040605} {\bibfield  {journal} {\bibinfo
  {journal} {Phys. Rev. Lett.}\ }\textbf {\bibinfo {volume} {124}},\ \bibinfo
  {pages} {040605} (\bibinfo {year} {2020})}\BibitemShut {NoStop}%
\bibitem [{\citenamefont {Ballentine}(2014)}]{Ballentine_2014_book}%
  \BibitemOpen
  \bibfield  {author} {\bibinfo {author} {\bibfnamefont {L}~\bibnamefont
  {Ballentine}},\ }\href@noop {} {\emph {\bibinfo {title} {Quantum Mechanics: A
  Modern Development}}}\ (\bibinfo  {publisher} {World scientific},\ \bibinfo
  {year} {2014})\BibitemShut {NoStop}%
\bibitem [{\citenamefont {Jermyn}(2005)}]{Jermyn_2005}%
  \BibitemOpen
  \bibfield  {author} {\bibinfo {author} {\bibfnamefont {Ian}\ \bibnamefont
  {Jermyn}},\ }\bibfield  {title} {\enquote {\bibinfo {title} {Invariant
  bayesian estimation on manifolds},}\ }\href {\doibase
  10.1214/009053604000001273} {\bibfield  {journal} {\bibinfo  {journal} {Ann.
  Statist.}\ }\textbf {\bibinfo {volume} {33}},\ \bibinfo {pages} {583--605}
  (\bibinfo {year} {2005})}\BibitemShut {NoStop}%
\bibitem [{\citenamefont {Amari}\ and\ \citenamefont
  {Nagaoka}(2007)}]{Amari2007}%
  \BibitemOpen
  \bibfield  {author} {\bibinfo {author} {\bibfnamefont {Shun-Ichi}\
  \bibnamefont {Amari}}\ and\ \bibinfo {author} {\bibfnamefont {Hiroshi}\
  \bibnamefont {Nagaoka}},\ }\href@noop {} {\emph {\bibinfo {title} {Methods of
  Information Geometry}}},\ Vol.\ \bibinfo {volume} {191}\ (\bibinfo
  {publisher} {American Mathematical Society},\ \bibinfo {year}
  {2007})\BibitemShut {NoStop}%
\bibitem [{\citenamefont {Amari}(1985)}]{Amari_1985}%
  \BibitemOpen
  \bibfield  {author} {\bibinfo {author} {\bibfnamefont {S.}~\bibnamefont
  {Amari}},\ }\href@noop {} {\emph {\bibinfo {title} {Differential-Geometrical
  Methods in Statistics}}}\ (\bibinfo  {publisher} {Springer-Verlag},\ \bibinfo
  {year} {1985})\BibitemShut {NoStop}%
\bibitem [{\citenamefont {Snoussi}(2007)}]{SNOUSSI2007163}%
  \BibitemOpen
  \bibfield  {author} {\bibinfo {author} {\bibfnamefont {Hichem}\ \bibnamefont
  {Snoussi}},\ }\bibfield  {title} {\enquote {\bibinfo {title} {Bayesian
  information geometry: Application to prior selection on statistical
  manifolds},}\ \ }(\bibinfo  {publisher} {Elsevier},\ \bibinfo {year} {2007})\
  pp.\ \bibinfo {pages} {163--207}\BibitemShut {NoStop}%
\bibitem [{\citenamefont {Jarzyna}\ and\ \citenamefont
  {Kołodyński}(2020)}]{Jarzyne_2020}%
  \BibitemOpen
  \bibfield  {author} {\bibinfo {author} {\bibfnamefont {Marcin}\ \bibnamefont
  {Jarzyna}}\ and\ \bibinfo {author} {\bibfnamefont {Jan}\ \bibnamefont
  {Kołodyński}},\ }\bibfield  {title} {\enquote {\bibinfo {title} {Geometric
  approach to quantum statistical inference},}\ }\href {\doibase
  10.1109/JSAIT.2020.3017469} {\bibfield  {journal} {\bibinfo  {journal} {IEEE
  Journal on Selected Areas in Information Theory}\ }\textbf {\bibinfo {volume}
  {1}},\ \bibinfo {pages} {367--386} (\bibinfo {year} {2020})}\BibitemShut
  {NoStop}%
\bibitem [{\citenamefont {Chentsov}(1978)}]{Chentsov_1978}%
  \BibitemOpen
  \bibfield  {author} {\bibinfo {author} {\bibfnamefont {N.~N.}\ \bibnamefont
  {Chentsov}},\ }\bibfield  {title} {\enquote {\bibinfo {title} {Algebraic
  foundation of mathematical statistics},}\ }\href@noop {} {\bibfield
  {journal} {\bibinfo  {journal} {Math. Operationsforsch. statist.}\ }\textbf
  {\bibinfo {volume} {9}},\ \bibinfo {pages} {267–276} (\bibinfo {year}
  {1978})}\BibitemShut {NoStop}%
\bibitem [{\citenamefont {Li}\ \emph {et~al.}(2018)\citenamefont {Li},
  \citenamefont {Pezzè}, \citenamefont {Gessner}, \citenamefont {Ren},
  \citenamefont {Li},\ and\ \citenamefont {Smerzi}}]{Yan_2018_entropy}%
  \BibitemOpen
  \bibfield  {author} {\bibinfo {author} {\bibfnamefont {Yan}\ \bibnamefont
  {Li}}, \bibinfo {author} {\bibfnamefont {Luca}\ \bibnamefont {Pezzè}},
  \bibinfo {author} {\bibfnamefont {Manuel}\ \bibnamefont {Gessner}}, \bibinfo
  {author} {\bibfnamefont {Zhihong}\ \bibnamefont {Ren}}, \bibinfo {author}
  {\bibfnamefont {Weidong}\ \bibnamefont {Li}}, \ and\ \bibinfo {author}
  {\bibfnamefont {Augusto}\ \bibnamefont {Smerzi}},\ }\bibfield  {title}
  {\enquote {\bibinfo {title} {Frequentist and bayesian quantum phase
  estimation},}\ }\href {\doibase 10.3390/e20090628} {\bibfield  {journal}
  {\bibinfo  {journal} {Entropy}\ }\textbf {\bibinfo {volume} {20}} (\bibinfo
  {year} {2018}),\ 10.3390/e20090628}\BibitemShut {NoStop}%
\bibitem [{\citenamefont {Bacharach}\ \emph {et~al.}(2019)\citenamefont
  {Bacharach}, \citenamefont {Fritsche}, \citenamefont {Orguner},\ and\
  \citenamefont {Chaumette}}]{Bacharach_2019_proceedinngs}%
  \BibitemOpen
  \bibfield  {author} {\bibinfo {author} {\bibfnamefont {Lucien}\ \bibnamefont
  {Bacharach}}, \bibinfo {author} {\bibfnamefont {Carsten}\ \bibnamefont
  {Fritsche}}, \bibinfo {author} {\bibfnamefont {Umut}\ \bibnamefont
  {Orguner}}, \ and\ \bibinfo {author} {\bibfnamefont {Eric}\ \bibnamefont
  {Chaumette}},\ }\bibfield  {title} {\enquote {\bibinfo {title} {A tighter
  bayesian cramÉr-rao bound},}\ }in\ \href {\doibase
  10.1109/ICASSP.2019.8683614} {\emph {\bibinfo {booktitle} {ICASSP 2019 - 2019
  IEEE International Conference on Acoustics, Speech and Signal Processing
  (ICASSP)}}}\ (\bibinfo {year} {2019})\ pp.\ \bibinfo {pages}
  {5277--5281}\BibitemShut {NoStop}%
\bibitem [{\citenamefont {{Caticha}}(2008)}]{Caticha_2008}%
  \BibitemOpen
  \bibfield  {author} {\bibinfo {author} {\bibfnamefont {Ariel}\ \bibnamefont
  {{Caticha}}},\ }\bibfield  {title} {\enquote {\bibinfo {title} {{Lectures on
  Probability, Entropy, and Statistical Physics}},}\ }\href@noop {} {\bibfield
  {journal} {\bibinfo  {journal} {arXiv e-prints}\ ,\ \bibinfo {eid}
  {arXiv:0808.0012}} (\bibinfo {year} {2008})},\ \Eprint
  {http://arxiv.org/abs/0808.0012} {arXiv:0808.0012 [physics.data-an]}
  \BibitemShut {NoStop}%
\bibitem [{\citenamefont {Correa}\ \emph {et~al.}(2015)\citenamefont {Correa},
  \citenamefont {Mehboudi}, \citenamefont {Adesso},\ and\ \citenamefont
  {Sanpera}}]{Correa_2015}%
  \BibitemOpen
  \bibfield  {author} {\bibinfo {author} {\bibfnamefont {Luis~A.}\ \bibnamefont
  {Correa}}, \bibinfo {author} {\bibfnamefont {Mohammad}\ \bibnamefont
  {Mehboudi}}, \bibinfo {author} {\bibfnamefont {Gerardo}\ \bibnamefont
  {Adesso}}, \ and\ \bibinfo {author} {\bibfnamefont {Anna}\ \bibnamefont
  {Sanpera}},\ }\bibfield  {title} {\enquote {\bibinfo {title} {Individual
  quantum probes for optimal thermometry},}\ }\href {\doibase
  10.1103/PhysRevLett.114.220405} {\bibfield  {journal} {\bibinfo  {journal}
  {Phys. Rev. Lett.}\ }\textbf {\bibinfo {volume} {114}},\ \bibinfo {pages}
  {220405} (\bibinfo {year} {2015})}\BibitemShut {NoStop}%
\bibitem [{\citenamefont {Ashida}\ \emph {et~al.}(2018)\citenamefont {Ashida},
  \citenamefont {Saito},\ and\ \citenamefont
  {Ueda}}]{Ashida_2018_thermalization}%
  \BibitemOpen
  \bibfield  {author} {\bibinfo {author} {\bibfnamefont {Yuto}\ \bibnamefont
  {Ashida}}, \bibinfo {author} {\bibfnamefont {Keiji}\ \bibnamefont {Saito}}, \
  and\ \bibinfo {author} {\bibfnamefont {Masahito}\ \bibnamefont {Ueda}},\
  }\bibfield  {title} {\enquote {\bibinfo {title} {Thermalization and heating
  dynamics in open generic many-body systems},}\ }\href {\doibase
  10.1103/PhysRevLett.121.170402} {\bibfield  {journal} {\bibinfo  {journal}
  {Phys. Rev. Lett.}\ }\textbf {\bibinfo {volume} {121}},\ \bibinfo {pages}
  {170402} (\bibinfo {year} {2018})}\BibitemShut {NoStop}%
\bibitem [{\citenamefont {Potts}\ \emph {et~al.}(2019)\citenamefont {Potts},
  \citenamefont {Brask},\ and\ \citenamefont
  {Brunner}}]{Potts_2019_fundamentallimits}%
  \BibitemOpen
  \bibfield  {author} {\bibinfo {author} {\bibfnamefont {Patrick~P.}\
  \bibnamefont {Potts}}, \bibinfo {author} {\bibfnamefont {Jonatan~Bohr}\
  \bibnamefont {Brask}}, \ and\ \bibinfo {author} {\bibfnamefont {Nicolas}\
  \bibnamefont {Brunner}},\ }\bibfield  {title} {\enquote {\bibinfo {title}
  {Fundamental limits on low-temperature quantum thermometry with finite
  resolution},}\ }\href {\doibase 10.22331/q-2019-07-09-161} {\bibfield
  {journal} {\bibinfo  {journal} {{Quantum}}\ }\textbf {\bibinfo {volume}
  {3}},\ \bibinfo {pages} {161} (\bibinfo {year} {2019})}\BibitemShut {NoStop}%
\bibitem [{\citenamefont {J\o{}rgensen}\ \emph {et~al.}(2020)\citenamefont
  {J\o{}rgensen}, \citenamefont {Potts}, \citenamefont {Paris},\ and\
  \citenamefont {Brask}}]{Jorgensen_2020_TightBound}%
  \BibitemOpen
  \bibfield  {author} {\bibinfo {author} {\bibfnamefont {Mathias~R.}\
  \bibnamefont {J\o{}rgensen}}, \bibinfo {author} {\bibfnamefont {Patrick~P.}\
  \bibnamefont {Potts}}, \bibinfo {author} {\bibfnamefont {Matteo G.~A.}\
  \bibnamefont {Paris}}, \ and\ \bibinfo {author} {\bibfnamefont {Jonatan~B.}\
  \bibnamefont {Brask}},\ }\bibfield  {title} {\enquote {\bibinfo {title}
  {Tight bound on finite-resolution quantum thermometry at low temperatures},}\
  }\href {\doibase 10.1103/PhysRevResearch.2.033394} {\bibfield  {journal}
  {\bibinfo  {journal} {Phys. Rev. Research}\ }\textbf {\bibinfo {volume}
  {2}},\ \bibinfo {pages} {033394} (\bibinfo {year} {2020})}\BibitemShut
  {NoStop}%
\bibitem [{\citenamefont {Paris}(2015)}]{Paris_2015}%
  \BibitemOpen
  \bibfield  {author} {\bibinfo {author} {\bibfnamefont {Matteo G~A}\
  \bibnamefont {Paris}},\ }\bibfield  {title} {\enquote {\bibinfo {title}
  {Achieving the landau bound to precision of quantum thermometry in systems
  with vanishing gap},}\ }\href {\doibase 10.1088/1751-8113/49/3/03lt02}
  {\bibfield  {journal} {\bibinfo  {journal} {Journal of Physics A:
  Mathematical and Theoretical}\ }\textbf {\bibinfo {volume} {49}},\ \bibinfo
  {pages} {03LT02} (\bibinfo {year} {2015})}\BibitemShut {NoStop}%
\bibitem [{\citenamefont {Mehboudi}\ \emph {et~al.}(2022)\citenamefont
  {Mehboudi}, \citenamefont {J\o{}rgensen}, \citenamefont {Seah}, \citenamefont
  {Brask}, \citenamefont {Ko\l{}ody\ifmmode~\acute{n}\else \'{n}\fi{}ski},\
  and\ \citenamefont {Perarnau-Llobet}}]{Mehboudi_accompany}%
  \BibitemOpen
  \bibfield  {author} {\bibinfo {author} {\bibfnamefont {Mohammad}\
  \bibnamefont {Mehboudi}}, \bibinfo {author} {\bibfnamefont {Mathias~R.}\
  \bibnamefont {J\o{}rgensen}}, \bibinfo {author} {\bibfnamefont {Stella}\
  \bibnamefont {Seah}}, \bibinfo {author} {\bibfnamefont {Jonatan~B.}\
  \bibnamefont {Brask}}, \bibinfo {author} {\bibfnamefont {Jan}\ \bibnamefont
  {Ko\l{}ody\ifmmode~\acute{n}\else \'{n}\fi{}ski}}, \ and\ \bibinfo {author}
  {\bibfnamefont {Mart\'{\i}}\ \bibnamefont {Perarnau-Llobet}},\ }\bibfield
  {title} {\enquote {\bibinfo {title} {Fundamental limits in bayesian
  thermometry and attainability via adaptive strategies},}\ }\href {\doibase
  10.1103/PhysRevLett.128.130502} {\bibfield  {journal} {\bibinfo  {journal}
  {Phys. Rev. Lett.}\ }\textbf {\bibinfo {volume} {128}},\ \bibinfo {pages}
  {130502} (\bibinfo {year} {2022})}\BibitemShut {NoStop}%
\end{thebibliography}%

\end{document}